\documentclass[apj]{emulateapj-rtx4}
\usepackage{apjfonts}
\usepackage{graphicx}
\usepackage{amssymb}
\usepackage{amsbsy}
\usepackage{amsmath}
\usepackage{mathrsfs}
\usepackage{color}
\usepackage[dvipdfmx]{hyperref}

\DeclareMathAlphabet{\mathbi}{OT1}{cmr}{bx}{it}
\SetMathAlphabet\mathbi{bold}{OT1}{cmr}{bx}{it}

\def\rd{{\rm d}}

\journalinfo{The Astrophysics Journal, 779, 110 (17pp), 2013, December 20}
\slugcomment{published 2013 November 27}

\begin{document}
\title{A Bayesian Approach to Estimate the Size and Structure of the Broad-line 
Region in Active Galactic Nuclei Using Reverberation Mapping Data}
\author{
  Yan-Rong Li\altaffilmark{1}, 
  Jian-Min Wang\altaffilmark{1,2}, 
  Luis C. Ho\altaffilmark{3,4}, 
  Pu Du\altaffilmark{1},
  and Jin-Ming Bai\altaffilmark{5, 6}
}
\altaffiltext{1}
{
Key Laboratory for Particle Astrophysics, Institute of High 
Energy Physics, Chinese Academy of Sciences, 19B Yuquan Road, 
Beijing 100049, China; liyanrong@mail.ihep.ac.cn
}

\altaffiltext{2}
{
National Astronomical Observatories of China, Chinese 
Academy of Sciences, 20A Datun Road, Beijing 100020, China
}

\altaffiltext{3}
{
Kavli Institute for Astronomy and Astrophysics, Peking University, Beijing
100871, China
}

\altaffiltext{4}
{
The Observatories of the Carnegie Institution for Science, 813 Santa 
Barbara Street, Pasadena, CA 91101, USA
}

\altaffiltext{5}
{
National Astronomical Observatories/Yunnan Observatory, 
Chinese Academy of Sciences, Kunming 650011, China
}

\altaffiltext{6}
{
Key Laboratory for the Structure and Evolution of Celestial Objects, 
Chinese Academy of Sciences, Kunming 650011, China
}

\begin{abstract}
This is the first paper in a series devoted to systematic study of the size 
and structure of the broad-line region (BLR) in active galactic nuclei (AGNs) 
using reverberation mapping (RM) data. We employ a recently developed Bayesian 
approach that statistically describes the variability as a damped random 
walk process and delineates the BLR structure using a flexible disk geometry 
that can account for a variety of shapes, including disks, rings, shells, and 
spheres.  We allow for the possibility that the line emission may respond 
non-linearly to the continuum, and we detrend the light curves when there is 
clear evidence for secular variation.  We use a Markov Chain Monte Carlo 
implementation based on Bayesian statistics to recover the parameters and 
uncertainties for the BLR model.  The corresponding transfer function is 
obtained self-consistently. We tentatively constrain the virial factor 
used to estimate black hole masses; more accurate determinations will have 
to await velocity-resolved RM data.  Application of our method to RM data
with H$\beta$ monitoring for about 40 objects shows that the assumed BLR geometry 
can reproduce quite well the observed emission-line fluxes from the continuum 
light curves. We find that the H$\beta$ BLR sizes obtained from our method are on 
average $\sim$20\% larger than those derived from the traditional cross-correlation method. 
Nevertheless, we still find a tight BLR size-luminosity relation with a slope 
of $\alpha=0.55\pm0.03$ and an intrinsic scatter of $\sim0.18$ dex.  In 
particular, we demonstrate that our approach yields appropriate BLR sizes for 
some objects (such as Mrk 142 and PG 2130+099) where traditional methods 
previously encountered difficulties.
\end{abstract}


\section{Introduction}
The well-established technique of reverberation mapping (RM) provides 
a  promising  pathway for directly measuring black hole mass in active 
galactic nuclei (AGNs) with broad emission lines (\citealt{Blandford1982, 
Peterson1993, Peterson2013}). Efforts over the past two decades have yielded 
RM measurements for $\sim50$ nearby Seyfert galaxies and quasars (e.g., 
\citealt{Bentz2013}) and led to the discovery of the widely used 
relationship between the size of the broad-line region (BLR) and the optical 
luminosity of the AGN (\citealt{Kaspi2000, Kaspi2005, Vestergaard2006, 
Bentz2009a, Bentz2013}), which serves as a cornerstone to study the demography of 
supermassive black holes in large AGN surveys (e.g., \citealt{GreeneHo2007, 
Shen2008, Vestergaard2009}) and secondary explorations of the 
role of supermassive black holes in various astrophysical contexts (e.g., 
\citealt{Marconi2004, Ho2008, Wang2009, Li2011, Li2012}).

The underlying principle of the RM technique is quite straightforward. Emission 
line variations are blurred echoes of continuum variations through the transfer 
function, which encodes the geometry and kinematic information of the BLR.
On both theoretical and observational side,
much attention has been paid to recovering the transfer function or velocity-delay 
map with the purpose of placing constraints on the basic properties and 
structure of the BLR (e.g., \citealt{Horne1991, Welsh1991, Wanders1995, Ulrich1996,
Kollatschny2003, Denney2009, Bentz2010,Goad2012,Grier2013b}). 
Traditional cross-correlation analysis, used in most previous reverberation
studies, has succeeded in characterizing BLR sizes using the derived time lags
(e.g., \citealt{Bentz2013} and references therein), although their interpretation 
in terms of realistic BLR structures remains elusive (\citealt{Netzer1990,Robinson1990,Welsh1999}). 
There are other more sophisticated mathematical methods developed for reconstructing 
the transfer function, including the maximum entropy technique (\citealt{Horne1994}), 
the regularized linear inverse method (\citealt{Krolik1995}), and the SOLA method 
(\citealt{Pijpers1994}). A physical model for the BLR is finally invoked to decode 
the transfer functions (e.g., \citealt{Horne2003, Bentz2010, Grier2013b}).

A major concern of these traditional methods is the assumption that the 
emission lines respond linearly to the ionizing continuum. This is the case for 
optically thin BLRs. However, for optically thick BLRs, while the total line and 
diffuse continua emission are proportional to the continuum flux, the emission
of individual lines might not. Photoionization calculations show that 
the responses of different emission lines depend on the ionization parameter
(e.g., \citealt{Netzer1985}). The situation is even more complicated when 
using the 5100~{\AA} continuum luminosity instead of the unobservable UV ionizing 
continuum for RM analysis. The shape of the incident continuum 
most likely does not remain constant during the reverberation variations.
Such evidence has been found in long-term RM monitoring of the well-studied Seyfert galaxy
NGC 5548 (\citealt{Dietrich1995, Peterson2002, Bentz2007}). Moreover, it is common for the variation 
amplitude of the emission lines to exceed that of the optical continuum (e.g., 
\citealt{Meusinger2011} and references therein). This is difficult to 
reconcile in the framework of linear response.  

On the other hand, due to our ignorance of the structure and kinematics of the BLR, 
a virial factor ($f_{\rm BLR}$) has to be assumed to convert the observed 
emission line widths and reverberation time lags into black hole mass. In the absence 
of any other direct black hole mass measurements, a common practice is to
calibrate $f_{\rm BLR}$ with the aid of the relationship between the black 
hole mass and stellar velocity dispersion of the bulge  of the host galaxy
($M_\bullet-\sigma_\star$ relation; \citealt{Onken2004, Woo2010, Park2013}), 
which is well established in local quiescent galaxies. The average value of 
$f_{\rm BLR}$ derived in the literature ranges from 
$\langle f_{\rm BLR}\rangle\approx 3$ (\citealt{Marconi2008, Graham2011}) up to 
$\sim 6$  (\citealt{Onken2004, Woo2010, Grier2013a}) with a scatter of about 0.4 dex, 
comparable to (\citealt{Gultekin2009}) or slightly larger than (Kormendy \& Ho 2013) 
that of the $M_\bullet-\sigma_\star$ relation.  The virial factor calibrated in this 
manner is valid only in a statistical sense. 
It seems likely that $f_{\rm BLR}$ differs from object to object 
(see the extensive discussion of \citealt{Goad2012}).
The dynamical timescale of the BLR, $R_{\rm BLR}/\Delta V$, is
on the order of a few years for typical values of $R_{\rm BLR}\approx 10$ light days
and $\Delta V\approx 2000~{\rm km~s^{-1}}$. Long-term monitoring of NGC 5548 
demonstrates that its BLR structure evolves year-to-year 
(\citealt{Wanders1996, Peterson2002, Shapovalova2004, Sergeev2007}), potentially
implying that $f_{\rm BLR}$ is also subject to variations. Furthermore, 
\cite{Pancoast2011} recently developed a fully general Bayesian framework to analyze 
RM data sets to model the geometry and dynamics of the BLR.  Subsequent application of this
technique on Mrk 50 yields a virial factor $f_{\rm BLR}\approx6$ (\citealt{Pancoast2012}), 
consistent with the nominal value commonly adopted, but for Arp 151 
the value of $f_{\rm BLR}\approx2.5$ seems somehow exceptional (\citealt{Brewer2011}). 
In this regard, an object-by-object determination of $f_{\rm BLR}$ will help 
to better understand the mass measurements and will permit a comprehensive exploration 
of BLR structures.

Following the Bayesian framework developed by \cite{Pancoast2011}, this paper 
is devoted to systematically study the structure of the BLR using RM data sets
with H$\beta$ monitoring accessible in the literature. Compared to the 
traditional methodology through the transfer function or velocity-delay maps, 
the Bayesian approach enables a direct probe of the BLR structure and a routine 
estimate of the modeling parameters. We take into account the non-linear response 
of emission lines to continuum variations, and we study the effect of detrending 
to remove secular variations that may contaminate the RM analysis. We attempt to 
recover the BLR geometry and place tentative constraints on the virial factors. 
We describe the methodology of our approach in 
Section 2 and the sample of RM light curves compiled from the literature
in Section 3. In Section 4, we present a general estimate to the 
virial factor for our assumed BLR geometry. Section 5 presents verifications of our 
approach, and our results are given in Section 6. We discuss 
the uncertainties of our method and future improvements in Section 7. 
The conclusions are summarized in Section 8.

Throughout the paper, we adopt a standard $\Lambda$CDM cosmology with
$\Omega_m=0.3$, $\Omega_{\rm \Lambda}=0.7$, and 
$H_0=72{\rm~km~s^{-1}~Mpc^{-1}}$. Unless stated otherwise, the BLR sizes derived 
from time series analysis always refer to rest-frame values (at redshift $z$); this 
and other variables invoking cosmic time 
are reduced by a factor of $(1+z)$ with respect to the values in the observer's 
frame. For the sake of brevity, 
when referring to the Julian Date, only the five least significant digits are retained.

\begin{figure*}[t!]
 \centering
 \includegraphics[angle=-90.0]{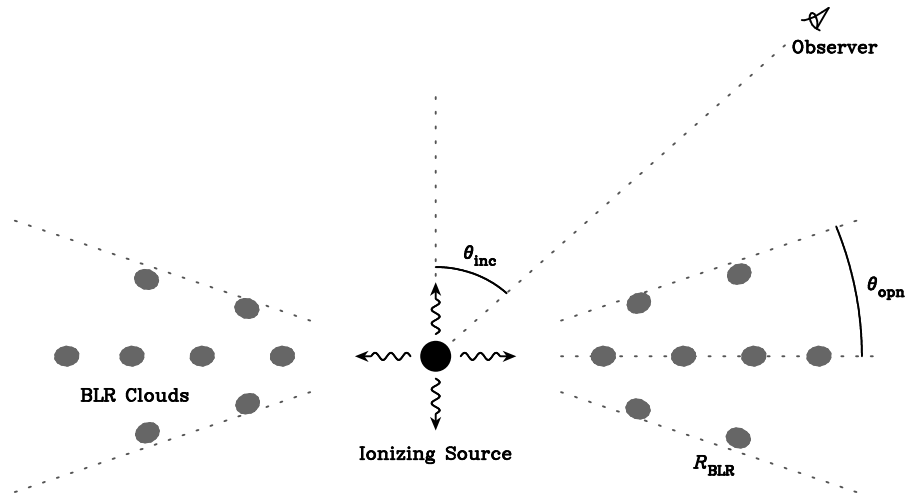}
 \caption{Schematic of the BLR geometry. The BLR has a flexible disk-like 
 geometry with an inclination angle
 $\theta_{\rm inc}$ to the observer and an opening angle $\theta_{\rm opn}$. 
 A central ionizing source is assumed to produce
 an isotropic UV emission that illuminates the surrounding BLR clouds (see Section 
 \ref{sec_modeling} for details).}
 \label{fig_sch}
\end{figure*}

\section{Methodology}
\subsection{Continuum Variability and Reconstruction}
\label{sec_continuum}
The first step in analysis of RM data is reconstructing the continuum 
light curve from usually irregularly sampled data.
For this purpose, we adopt and slightly modify the framework outlined by 
\cite{Rybicki1992}.  For the sake of comparison, the notation used here also 
follows \cite{Rybicki1992}.  Detailed derivations of the following equations 
are given in Appendix \ref{app_eqn}.

Let the column vector $\mathbi{y}$ denote a set of $m$ measurements for 
a light curve in a monitoring campaign. In practice, each measurement 
$y_i$ can be deemed to be the sum of an underlying signal $s_i$ 
representing the variation, a constant $q$ representing the mean of 
the light curve, and a noise value $n_i$ representing the associated 
measurement error. Written in a concise form of column vectors,
\begin{equation}
 \mathbi{y} = \mathbi{s+n}+\mathbi{E}q,
\end{equation}
where $\mathbi{E}$ is a vector with all unity elements (i.e. $E_i=1$).
We will demonstrate below that it is particularly necessary to
separate out the mean before performing the reconstruction.

Without any independent statistical information about the signal $\mathbi{s}$,
a practical strategy is to assume that $\mathbi{s}$ is stationary. This 
simplifies the covariance function of $\mathbi{s}$, denoted by $S(t_1, t_2)$, 
between times $t_1$ and $t_2$ in a way that $S(t_1, t_2)$ depends only on the 
time difference $t_1-t_2$ (\citealt{Rybicki1992}). Recent work by
\cite{Kelly2009} finds that the covariance function driven by a damped 
random walk model, expressed as
\begin{equation}
S(t_1, t_2)=\sigma_{\rm d}^2\exp\left[-\left(\frac{|t_1-t_2|}{\tau_{\rm d}}\right)^{\alpha}\right],
\label{eqn_cov}
\end{equation}
can well describe the optical variability of AGNs. This is further reinforced 
by subsequent investigations of large samples of AGN light curves 
(\citealt{Kozlowski2010, MacLeod2010, MacLeod2011, Kelly2011, Zu2013}). Here, 
$\tau_{\rm d}$ is the typical timescale of variation, $\sigma_{\rm d}$ is 
the standard deviation of variation on long-timescale ($\gg\tau_{\rm d}$), and 
$\alpha$ is a smoothness parameter.  Previous studies show that $\alpha=1$ is 
sufficient for interpreting the variabilities (\citealt{MacLeod2010, Zu2013}),
and we therefore fix it throughout the calculations.

If we further assume that both $\mathbi{s}$ and $\mathbi{n}$ are Gaussian and 
uncorrelated, the probability for a realization of $\mathbi y$ is 
(see Appendix \ref{app_eqn} for details)
\begin{equation}
P(\mathbi{y}|\sigma_{\rm d}, \tau_{\rm d})=\frac{1}{\sqrt{(2\pi)^m|\mathbi{C}|}}
\exp\left\{-\frac{1}{2}\left(\mathbi{y}-\mathbi{E}\hat q\right)^T\mathbi{C}^{-1}
\left(\mathbi{y}-\mathbi{E}\hat q\right)\right\},
\end{equation}
where superscript ``$T$'' denotes the transposition, $\mathbi{C\equiv S+N}$, $\mathbi{S}$ is
the covariance matrix of signal $\mathbi{s}$ 
given by Equation (\ref{eqn_cov}), $\mathbi{N}$ is the covariance matrix
of the noise $\mathbi{n}$, and the best estimate of $q$ is (see also 
\citealt{Rybicki1992})
\begin{equation}
\hat q=\frac{\mathbi{E}^T\mathbi{C}^{-1}\mathbi{y}}{\mathbi{E}^T\mathbi{C}^{-1}\mathbi{E}}.
\end{equation}

Now, given a set of measurement, we recover the damped 
random walk process with the aid of Bayes' theorem:
\begin{eqnarray}
&&P(\sigma_{\rm d}, \tau_d|\mathbi{y})=
\frac{P(\sigma_{\rm d}, \tau_{\rm d})P(\mathbi{y|\sigma_{\rm d}, \tau_{\rm d}})}{P(\mathbi{y})}\nonumber\\
&=&\frac{P(\sigma_{\rm d}, \tau_{\rm d})}{\sqrt{(2\pi)^m|\mathbi{C}|}}
\exp\left\{-\frac{1}{2}\left(\mathbi{y}-\mathbi{E}\hat q\right)^T\mathbi{C}^{-1}
\left(\mathbi{y}-\mathbi{E}\hat q\right)\right\},
\label{eqn_post_var}
\end{eqnarray}
where the marginal likelihood $P(\mathbi{y})$ is merely a normalization factor
that is neglected (\citealt{Sivia2006}). Maximizing this posterior distribution
yields best estimates for $\sigma_{\rm d}$ and $\tau_{\rm d}$. Specifically, we assign 
logarithmic priors for $\sigma_{\rm d}$ and $\tau_{\rm d}$ in Equation 
(\ref{eqn_post_var}), first employ the simulated annealing algorithm
(\citealt{Liu2001}) to locate the solutions maximizing Equation (\ref{eqn_post_var}),
and then enter a Markov chain Monte Carlo analysis to explore the statistical properties 
of $\sigma_{\rm d}$ and $\tau_{\rm d}$.

After determining the best values for $\sigma_{\rm d}$ and $\tau_{\rm d}$, 
the most probable estimate of the light curve at any time $t_\star$ 
is (see Appendix \ref{app_eqn})
\begin{equation}
\hat y_\star=\mathbi{S}_\star^T\mathbi{C}^{-1}(\mathbi{y}-\mathbi{E}\hat q)+\hat q,
\label{eqn_est}
\end{equation} 
where $\mathbi{S}_\star$ is a vector of the covariances between $t_\star$ and 
the time of each measurement point (given by Equation (\ref{eqn_cov})).
The mean square residual of this estimate is
\begin{equation}
\langle\Delta \hat y_\star^2\rangle
=\sigma_{\rm d}^2 - \mathbi{S}_\star^T\mathbi{C}^{-1}\mathbi{S}_\star
+\frac{\left(\mathbi{S}_\star^T\mathbi{C}^{-1}\mathbi{E}-1\right)^2}
{\mathbi{E}^T\mathbi{C}^{-1}\mathbi{E}}.
\end{equation}
From Equation (\ref{eqn_cov}), we note that at gaps far from any data points 
the covariance matrix $\mathbi{S}_\star$ approaches zero; thus, the estimate 
given by Equation (\ref{eqn_est}) tends toward the mean of the light curve with its 
uncertainty increasing up to the standard deviation of the variation ($\sigma_{\rm d}$). 
Meanwhile, if the mean $q$ of the light curve is not separated out
as in Equation (\ref{eqn_est}), the best estimate $\hat y$ tends toward zero at data gaps
and hence give rise to a bias (see discussion in \citealt{Rybicki1992}).

As shown by \cite{Rybicki1992}, a typical reconstruction of the observed light curve
is obtained by adding to the most probable estimate $\hat y_\star$ a 
Gaussian random process with zero mean and covariance matrix 
$(\mathbi{S}^{-1}+\mathbi{N}^{-1})^{-1}$.

It is worth mentioning that in the damped random walk model the variation of the
light curve on short timescales ($t\ll\tau_{\rm d}$) is 
$\sim\sigma_{\rm d}\sqrt{2t/\tau_{\rm d}}$ (\citealt{Kelly2009}). Therefore, 
one can roughly obtain an order-of-magnitude estimate of $\tau_{\rm d}$ by 
simple inspection of the amount of variation. Larger $\tau_{\rm d}$ leads to 
slower variations and smoother light curves. Also, the parameters $\tau_{\rm d}$ 
and $\sigma$ are found to be correlated with the physical properties of accretion 
disks, including optical luminosities, Eddington ratios, and black hole masses 
(\citealt{Kelly2009, MacLeod2011}). \cite{Dexter2011} demonstrate the ability 
of the damped random walk model for accretion disk fluctuations to explain the 
microlensing observations of accretion disks.

\subsection{Broad-line Region Modeling}
\label{sec_modeling}
A number of lines of evidence suggest that the BLR has a flattened shape: (1)
emission lines with disk-like (e.g., double-peaked) profiles are common 
in AGNs (e.g., \citealt{Eracleous2003, Strateva2003, Gezari2007,Lewis2010} and 
references therein); (2) Balmer line widths correlate with the orientation of 
the rotation axis of the BLR (e.g., \citealt{Rokaki2003, Jarvis2006, 
Decarli2008}); (3) possible detection of orbital motion of the BLR 
(e.g., \citealt{Sergeev2000, Eracleous2009}).  Furthermore, a number of 
previous velocity-resolved RM studies show the BLR to be disk-like (e.g., 
\citealt{Bentz2010, Grier2013b, Peterson2013}).  Motivated by these 
results, it is reasonable at this stage to assume that the BLR has a 
disk-like structure. We follow \cite{Pancoast2011}'s approach, which has been 
successfully applied to the RM data of Arp 151 (\citealt{Brewer2011}) and 
Mrk 50 (\citealt{Pancoast2012}). The details of the modeling can be found in
these original works.  We here describe the essential points and our 
improvements thereof.

Figure 1 shows a schematic of the BLR geometry adopted in the present work.
The BLR is represented by a large number of discrete, point-like clouds, which 
orbit around 
the central black hole and absorb the ionizing continuum from the central 
source (i.e. accretion disk) and re-radiate emission lines. The distribution of 
the clouds is a flexible axisymmetric disk, which, with suitable parameters, can account 
for a variety of shapes, including shells, spheres, and rings. 
The radial distribution of clouds is parameterized according to
\begin{equation}
R=F\mu+(1-F)\mathscr{R},
\label{eqn_cloud}
\end{equation}
where $\mathscr{R}$ is a random variable drawn from a Gamma distribution with 
a mean $\mu$ and a standard deviation $\beta\mu$. In this configuration, 
the overall mean radius for the cloud distribution is $\mu$, the inner hard edge of the BLR
is $R_{\rm in}=F\mu$, and the mean radial width of the BLR is $\sigma_R\approx\mu\beta(1-F)$. 
The parameter $\beta$ controls the shape of the cloud distribution: small values 
of $\beta$ create narrow normal distributions, while large values tend to create 
exponential distributions. The value of $F$ lies in a range $0-1$. According to 
the previous works of \cite{Brewer2011} and \cite{Pancoast2012}, it is adequate 
to set the range of $\beta$ to $0-1$.

The BLR clouds subtend a solid angle denoted by an opening angle $\theta_{\rm opn}$,
which is defined so that $\theta_{\rm opn}$ approaching zero creates thin disks/rings and
$\theta_{\rm opn}$ approaching $\pi/2$ creates spheres/shells. Within the opening angle,
we assume that the clouds are uniformly distributed over the polar and azimuthal directions.
Lastly, the BLR is viewed at an inclination angle $\theta_{\rm inc}$ to the distant 
observer, which is defined so that an inclination of zero corresponds to face-on, and
an inclination of $\pi/2$ corresponds to edge-on. We give some specific cases for 
illustration purposes: For $\theta_{\rm opn}=90^\circ$, one obtains a thin spherical 
shell with $F\rightarrow1$ and $\beta\rightarrow0$ and a thick sphere
with $F\rightarrow0$ and $\beta\rightarrow1$. 

Before producing the observed emission-line fluxes, we list a necessary 
{\em ansatz}: (1) The UV ionizing continuum is simply proportional to the optical
{5100 \AA} continuum; however, we relax the usual assumption of linear response of 
the emission lines (see also the early work by \citealt{Gaskell1986}). 
(2) The ionizing source has a point-like geometry so that its emission
is isotropic and falls off with the square of the distance.  (3) All BLR clouds have the 
same size and density, and there is no shadowing among clouds. Given a cloud 
distribution, we predict the emission-line flux response to the continuum at 
time $t$ by summing over the emission from all the clouds:
\begin{equation}
f_{l}(t)=\sum_i\epsilon_i(t)=A\sum_{i}w_i\left[I_i\frac{f_c(t-\tau_i)}{R_i^2}\right]^{1+\gamma},
\label{eqn_line}
\end{equation}
where $\tau_i$ is the time lag of the re-radiation from the $i$th cloud at 
distance $R_i$ to the central source, $A$ is a response coefficient, $w_i$ is 
the weight of the cloud in response to the continuum, $I_i$ is a flexible parameter 
describing any possible anisotropic effects and deviations from the inverse 
square decline of the continuum flux, and $\gamma$ denotes the non-linearity of the response. 
The uncertainties inherent from measurement errors of the continuum are 
\begin{equation}
 \Delta f_{l}(t) \approx (1+\gamma)A\sum_{i}w_i\left(\frac{I_i}{R_i^2}\right)^{1+\gamma}
 f_c^\gamma(t-\tau_i)\Delta f_c(t-\tau_i), 
\label{eqn_err}
\end{equation}
where the variation amplitude is assumed to be small (as in most cases) and a linear 
expansion is used. 

Rewriting Equations (\ref{eqn_line}) and (\ref{eqn_err}) into a general form,
\begin{equation}
f_{l}(t)=A\int \Psi(\tau)f_c^{1+\gamma}(t-\tau)\rd\tau,
\label{eqn_rm}
\end{equation}
and 
\begin{equation}
\Delta f_{l}(t) =  (1+\gamma)A\int\Psi(\tau)f_c^\gamma(t-\tau)\Delta f_c(t-\tau)\rd\tau,
\end{equation}
with the transfer function 
\begin{equation}
\Psi(\tau)=\sum_i\delta(\tau-\tau_i)w_i\left(\frac{I_i}{R_i^2}\right)^{1+\gamma},
\end{equation}
where $\delta(x)$ is the Dirac delta function.
If we let $\gamma=0$ in Equation (\ref{eqn_rm}), one arrives at the traditional 
linear response. 

At this stage, we focus on the variation of the light curves, and the absolute 
units of the light curves are no longer important; thus, the coefficient $A$ is a nuisance 
parameter. For simplicity, we also fix the weights to $w_i=1$ and neglect $I_i$ throughout 
the calculations. We only make use of the velocity-unresolved RM data sets such that 
the velocity information of the clouds disappears in Equation (\ref{eqn_rm});
however, this can be readily added (see \citealt{Brewer2011, Pancoast2011, 
Pancoast2012}). 

The role of the free parameters in shaping the light curves of the emission lines are as follows.
The mean radius $\mu$ determines the overall time lags of the emission lines. 
Larger inclination $\theta_{\rm inc}$ tends to broaden the transfer functions toward
double peaks, while larger opening angle $\theta_{\rm opn}$ tends to broaden the 
transfer functions toward top-hat. The parameters $\beta$ and $F$ jointly control 
the width of the BLR and therefore the width of the transfer function. In addition,
$F$ sets a hard inner limit on the BLR. The parameter $\gamma$ determines the amplitude 
of response to the continuum variations. It is apparent that 
the parameters $\theta_{\rm inc}$ and $\theta_{\rm opn}$ are degenerate at some level, as are 
the parameters $\beta$ and $F$. In general, there is also degeneracy among
other parameters, depending on the quality of the RM data. High-fidelity data sets,
namely those with a fine sampling rate, higher signal-to-noise ratio, and 
velocity-resolved information,  are beneficial to eliminate these degeneracies.

The terminology ``BLR size'' needs to be clarified a bit here.
We introduce two definitions: mass-weighted average radius
\begin{equation}
R_m=\frac{\sum_i m_i R_i}{\sum m_i},
\end{equation}
where $m_i$ is the mass of $i$th cloud, and emissivity-weighted average radius
\begin{equation}
R_e=\frac{\sum_i\epsilon_i R_i}{\sum_i \epsilon_i}.
\end{equation}
Because BLR clouds are treated as point masses with the same density and size,  
$R_m$ is exactly equal to the mean radius of the cloud distribution.
From Equation (\ref{eqn_line}), it is apparent that the emissivity of each 
cloud is variable with time in response to the continuum variation, and hence
the emissivity-averaged radius is weakly time-dependent%
\footnote{Note that BLRs may undergo secular evolution, leading to
mass/emissivity-averaged radius varying significantly on a timescale of years
(e.g., in NGC 5548; \citealt{Wanders1996, Peterson2002}).}.
It is convenient to use the time-averaged value for $R_e$ over the duration of 
the RM campaign. It is easy to verify that generally $R_e\leqslant R_m$ since 
the ionization flux most likely declines with radius. However, in our present modeling via
Equation (\ref{eqn_cloud}), the clouds are mostly distributed around the mean radius $\mu$,
leading to $R_e\approx R_m$ in most cases. We therefore hereafter only use $R_m$ to
refer to the BLR size.

\subsection{Detrending of Light Curves}
As shown by \cite{Welsh1999}, long-term secular variability is incidentally
detected over the duration of the campaign, which is found to be uncorrelated 
with reverberation variations (\citealt{Sergeev2007}) and thus will bias the 
desired correlation analysis between the continuum and emission light curves.
The continuum and emission lines may display different secular trends.
Plausible causes are, but not limited to, the non-linear response mentioned 
in the previous section, variations in the shape of the ionization continuum, 
as well as secular evolution of BLR structure, which is independent of the 
ionization sources.

To remove the bias due to secular variability, we adopt a first-order polynomial to 
fit the light curve with its mean subtracted and then detrend the original 
light curve by removing the polynomial. The goal of subtracting the mean 
before fitting is to keep the mean of the light curve unchanged. Detrending is 
necessary only when there is evident secular variability, such that the continuum
and the emission light curves exhibit different secular trends. In such cases, 
the results of fitting the predicted emission line fluxes to observations
will be substantially improved after detrending.  In practice, we apply the 
detrending when the fitting to the emission lines is judged to be poor, which 
we define as $\chi^2/{\rm dof}>1.5$ (see below for the definition of $\chi^2$).
The choice of this limit is a bit arbitrary, but it suffices to identify 
potential cases for detrending. The detrending is deemed acceptable if $\chi^2/{\rm dof}$ diminishes.

\subsection{Markov Chain Monte Carlo Implementation}
The measured RM data in hand are the line flux time series with associated 
errors $(y_l, \sigma_l)$ and the continuum flux time series with associated 
errors $(y_c, \sigma_c)$; in most cases, both are irregularly sampled. 
We first reconstruct the continuum time series 
using the damped random walk model described in Section \ref{sec_continuum}.
Given the geometry model of the BLR with parameter set $\boldsymbol \theta$,  
a reconstruction of the line flux from the continuum flux series can be made 
as described in the preceding section; we denote the reconstructed flux by 
$y_{p}(y_c|\boldsymbol\theta)$ with errors $\sigma_p$. Suppose that the probability distribution for 
the measurement errors are Gaussian and uncorrelated.  The likelihood function 
can then be written as
\begin{equation}
P(D|\boldsymbol\theta)=\prod_{i=1}^{m} \frac{1}{\sqrt{2\pi(\sigma_l^2+\sigma_p^2)}}
\exp\left(-\frac{1}{2}\frac{\left[y_l-y_p(y_c|\boldsymbol\theta)\right]^2}{\sigma_l^2+\sigma_p^2}\right),
\label{eqn_post}
\end{equation} 
where $D$ represents the measured data. Again, from Bayes' theorem, the posterior 
probability distribution for $\boldsymbol\theta$ is given by
\begin{equation}
 P(\boldsymbol\theta|D)=\frac{P(\boldsymbol\theta)P(D|\boldsymbol\theta)}{P(D)},
 \label{eqn_prob}
\end{equation}
where the marginal likelihood $P(D)$ is a normalization factor that is irrelevant to the
subsequent analysis. The prior probabilities $P(\boldsymbol\theta)$ in 
Equation (\ref{eqn_prob}) are assigned as follows: for parameters whose 
typical value ranges are known, a uniform prior is assigned; otherwise, if the 
parameter information is completely unknown, a logarithmic prior is assigned 
(\citealt{Sivia2006}). Among the seven free parameters, the priors for the mean BLR 
size $\mu$ and the response coefficient $A$ are set to be logarithmic, and the 
rest (inclination $\theta_{\rm opn}$, opening angle $\theta_{\rm opn}$, non-linearity parameter $\gamma$, 
and radial distribution parameters $\beta$ and $F$) are set to uniform.

We employ the Markov Chain Monte Carlo method with parallel tempering and the
Metropolis-Hastings algorithm (e.g., \citealt{Liu2001}) to construct samples 
from the posterior probability distribution, and then explore the statistical 
properties of the model parameters. The parallel tempering algorithm guards the Markov 
chain against being stuck in a local maximum and expedites its convergence to globally 
optimized solutions. Depending on the individual object, the free parameters are probably 
correlated with each other. Using the covariance matrix of free parameters will improve
the efficiency of Monte Carlo sampling. We initially input a diagonal covariance matrix 
and recompute it every 10,000 steps based on the newly generated section of the Markov chain.
Empirically, after one or two iterations, the correlation matrix turns out to be stable
and the Markov chain rapidly converges. The Markov chain is run 50,000 steps in total
with $10^6$ BLR clouds. Unless stated otherwise, the best estimates for the parameters are 
taken to be the expectation value of their distribution and the uncertainties
are taken to be the standard deviation. Presently we are unable to evaluate the 
systematic uncertainties inherent in the models.

We also introduce a $\chi^2$ for the fitting of the emission lines 
according to Equation (\ref{eqn_post}):
\begin{equation}
\chi^2 = \sum_{i=1}^m\frac{\left[y_l - y_p(y_c|\boldsymbol{\theta})\right]^2}{\sigma_l^2 + \sigma_p^2}.
\end{equation}
There are seven free parameters for BLR modelings. Accordingly, the reduced $\chi^2$ 
is calculated from $m-7$ degrees of freedom (dof) for light curves with $m$ observations.
As described in the preceding section, $\chi^2$/dof is used to determine whether detrending is necessary or acceptable.

\section{Data Sample}
We extract light curves of all objects with H$\beta$ monitoring 
accessible in the literature to date%
\footnote{The International AGN Watch project
provides machine-readable data tables for several objects,
which can be directly downloaded from its Web site 
(http://www.astronomy.ohio-state.edu/$\sim$agnwatch).},
mostly from the homogeneous compilations of \cite{Peterson2004} and 
\cite{Bentz2009a, Bentz2013}. 
We make use of the data that were designated by \cite{Bentz2013}
as reliable (see their Table 13).
The properties of all objects are summarized in Table \ref{tab_obj}.
\cite{Bentz2009a, Bentz2013} used surface brightness
decomposition of {\it Hubble Space Telescope}\ images to measure the 
starlight contribution of the host galaxies to the optical luminosity 
of the central nuclei for most of the RM objects.
We use their host-corrected 
luminosities in order to redetermine the $R_{\rm BLR}-L$ relationship. The Lick AGN 
Monitoring Project (LAMP) published photometric $B$-band and $V$-band continuum 
light curves (\citealt{Bentz2009b, Walsh2009}); we use their $V$-band 
photometric light curves. The magnitudes are converted 
into fluxes and then normalized to unity for computational convenience.
The following objects require special treatment or comments.

\begin{enumerate}
 \item For Fairall 9, the sampling rate at 1390 {\AA} was much higher than
that in the optical band.  Since the two bands vary
simultaneously (\citealt{Rodriguez1997,Santos1997}), we use the UV data at 
1390 {\AA} as a surrogate for the more poorly sampled optical data.

 \item NGC 3227 is known to be one of the most heavily reddened objects in the RM 
sample.  Bentz et al. (2013) corrected its optical luminosity at 5100 {\AA} by 
an extinction of 0.26 dex based on the reddening curve of \cite{Crenshaw2001}. 

 \item 3C 120 was quite poorly sampled during the 1989--1996 campaign.
Although the monitoring duration lasted as eight years, in total there were 
only 52 observations, with time gaps of about 100--300 days.
To alleviate possible aliasing effects, we discard the beginning and 
ending parts of the light curves with particularly sparse sampling
and only use the data between 1992 and 1996 (JD48869--JD50101).

\item The monitoring campaign of NGC 3783 was undertaken by the International AGN Watch 
(\citealt{Stirpe1994}). \cite{Onken2002} recalibrated the optical spectra based on a refined 
algorithm. We use their revised data set for analysis.

\item NGC 7469 was monitored at MDM Observatory during 2010--2011 
(\citealt{Grier2012}) but is missing from Table \ref{tab_obj} because these
data are not yet publicly accessible.
\end{enumerate}

\cite{Peterson2004} reanalyzed the broad emission lines of all RM data available
at that time and compiled line widths derived from the rms spectra. We directly 
use their measurements of H$\beta$ line width for all the objects 
included in their compilation. For other objects, we quote line widths derived 
similarly from rms spectra given by the corresponding references listed in Table \ref{tab_obj}.
%

%
\begin{figure}[t!]
\centering
\includegraphics[angle=-90.0, width=0.45\textwidth]{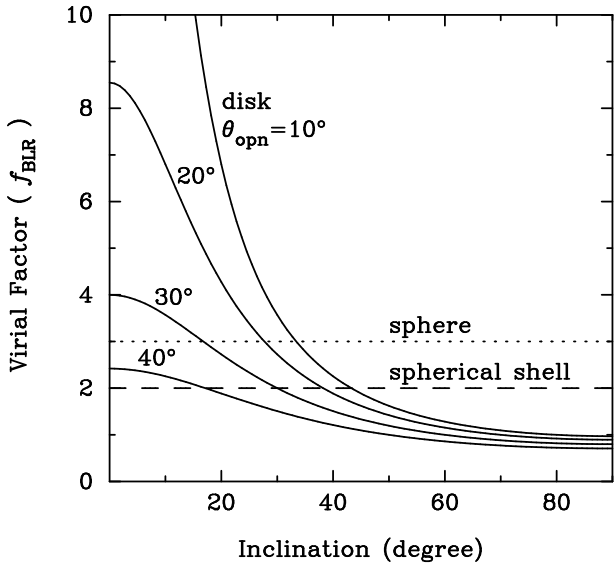}
\caption{Dependence of the virial factor on inclination angle for different opening angles; 
see Equation (\ref{eqn_factor}). Dashed and dotted lines represent the virial factors for a spherical shell
and sphere, respectively.}
\label{fig_factor}
\end{figure}

\begin{figure*}[t!]
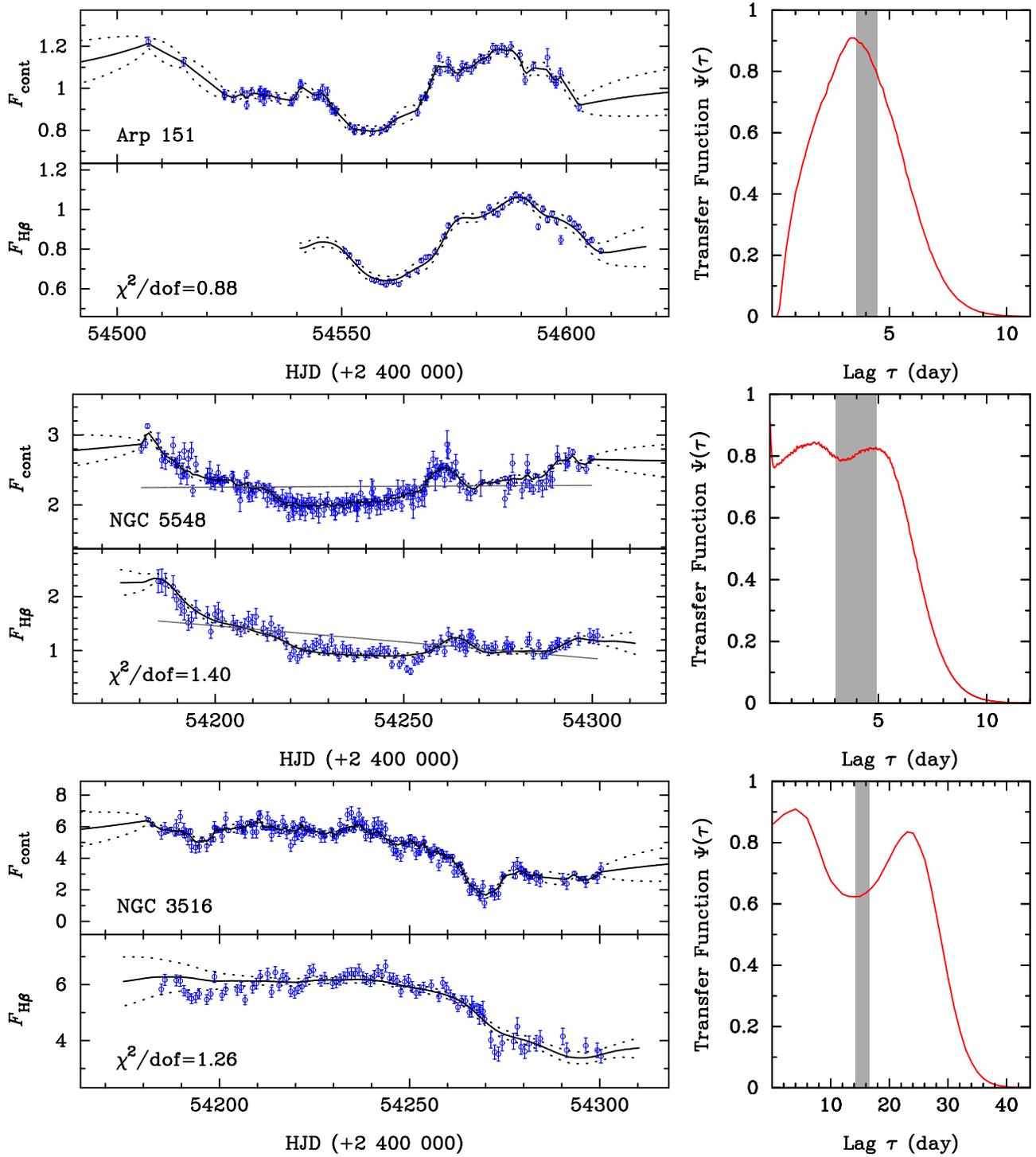

\centering
\includegraphics[angle=-90.0, width=0.95\textwidth]{arp151.ps}
\includegraphics[angle=-90.0, width=0.95\textwidth]{ngc5548.ps}
\includegraphics[angle=-90.0, width=0.95\textwidth]{ngc3516.ps}
\caption{Three illustration cases: Arp 151, NGC 5548, and NGC 3516. 
({\em Left}) Data points with error bars are the observed light curves. Thick 
solid lines show the reconstructed continuum at 5100 \AA\ (top panel) and the 
recovered H$\beta$ emission (bottom panel); dashed lines represent the 
uncertainties.  For NGC 5548 (middle panels), the gray line shows the fit for 
the detrending applied to remove secular variation.  ({\em Right}) Transfer 
function $\Psi(\tau)$ (red curve, in arbitrary units) derived from the best-fit
BLR model. The vertical shaded area represents the recovered BLR size and its 
uncertainty.}
\label{fig_cases}
\end{figure*}

%
%
\section{The Virial Factor}
Assuming that the motion of BLR clouds is dominated by the gravity of the central black hole, 
the "virial" black hole mass can be obtained trivially through RM:
\begin{equation}
M_{\rm vir}= f_{\rm BLR}\frac{V_{\rm obs}^2R_{\rm BLR}}{G},
\label{eqn_mvir}
\end{equation}
where $G$ is the gravitational constant and $V_{\rm obs}$ is the emission-line velocity width.
Here, the dimensionless so-called virial factor $f_{\rm BLR}$ subsumes 
all the unknown properties of the BLR, including its geometry, kinematics, and 
inclination angle. In the present approach, the BLR geometry is recovered statistically, 
thus allowing the virial factor to be estimated in an approximate sense
(since the kinematics remains unknown at this stage).

For a thick Keplerian disk, the observed line width is related to 
the Keplerian velocity by (e.g., \citealt{Collin2006}) 
\begin{equation}
V_{\rm obs}\approx V_{\rm Kep}\left(h^2+\sin^2\theta_{\rm inc}\right)^{1/2},
\label{eqn_width}
\end{equation}
where $h$ is the aspect ratio of the disk and $\theta_{\rm inc}$ is the 
inclination angle.  The aspect ratio is expressed in terms of the opening 
angle of the disk as $h\approx\sin\theta_{\rm opn}$. Bear in mind that the 
line width measurement depends on the line profile and that there might be a 
scale factor of order of unity in Equation 
(\ref{eqn_width}) (e.g., \citealt{Collin2006}). Combining Equation (\ref{eqn_mvir}) with
the true black hole mass estimate 
\begin{equation}
M_\bullet=\frac{V_{\rm Kep}^2R_{\rm BLR}}{G},
\end{equation}
the viral factor becomes
\begin{equation}
 f_{\rm BLR}\approx\left(\sin^2\theta_{\rm opn}+\sin^2\theta_{\rm inc}\right)^{-1}.
\label{eqn_factor}
\end{equation}

Note that the above equation is valid only for a disk-like structure. When the opening angle
approaches $90^\circ$, the BLRs become a spherical shell or a sphere. In such 
a case, 
if we further assume that the cloud motions are isotropic, the virial factor will 
be $f_{\rm BLR}\approx 2$ for thin shells and $f_{\rm BLR}\approx 3$ for 
spheres (\citealt{Netzer1990}),  independent of the inclination angle. Figure \ref{fig_factor} 
illustrates the variation of the virial factor with inclination for different opening angles. 
As expected, the virial factor is highly sensitive to inclination for small opening angles and 
decreases with opening angle for a given inclination. In the following, we estimate 
$f_{\rm BLR}$ using Equation (\ref{eqn_factor}) if $\theta_{\rm opn}<40^\circ$; otherwise, we
set $f_{\rm BLR}\approx 2-3$. Here $V_{\rm obs}$ refers to the velocity dispersion of the
emission line, $\sigma_{\rm line}$.  If line width is parameterized instead by FWHM, the corresponding 
$f_{\rm BLR}$ drops roughly by a factor of four.

%
\section{Verifications of Our Approach}
\subsection{An Illustration Case: Arp 151}
This object has been intensively studied in previous works based on the same RM data from the  
LAMP project (\citealt{Bentz2009b, Walsh2009}). \cite{Bentz2010} recovered the velocity-delay 
maps for multiple emission lines using the maximum-entropy technique and concluded that, 
although the constraints are not definitive, a plausible warped disk-like BLR is preferred.
\cite{Brewer2011} used a Bayesian framework for velocity-resolved RM to 
directly measure the central black hole mass. Their results clearly suggest that a disk-like 
structure for the BLR can reproduce the observations quite well. They derived a virial factor 
$f_{\rm BLR}=2.5\pm1.6$ and a black hole mass $M_\bullet = (3.2\pm2.1)\times10^6\,M_\odot$.

For the purpose of comparison, we plot our reconstruction of the continuum and H$\beta$ line 
fluxes in the top panels of Figure \ref{fig_cases}. The typical rest-frame damping timescale 
for the damped random walk process is $\log(\tau_{\rm d}/{\rm day})=1.88\pm0.59$.
The $\chi^2$/dof of the fit to the H$\beta$ line flux series is 0.88. 
The top right panel of Figure \ref{fig_cases} shows the recovered transfer function, which is
remarkably consistent with the results from the maximum-entropy technique (see Figure 1 
of \citealt{Bentz2010}). The best-fit values for the BLR parameters are $R_{\rm d}=4.0\pm0.4$, 
$\theta_{\rm inc}=29^\circ\pm18^\circ$, and 
$\theta_{\rm opn}=36^\circ\pm22^\circ$; these agree well with the model of \cite{Brewer2011}%
{\footnote{Note that \cite{Brewer2011} defined the inclination angle to be the 
complement of the usual convention adopted in the present calculations, and 
their opening angle is twice of that of our definition.}.
According to Figure \ref{fig_factor}, this gives a virial factor $f_{\rm BLR}\approx1.8$, which
implies that the simple estimate given by Equation (\ref{eqn_factor})
is quite acceptable and that, provided with high-quality RM data,  even velocity-unresolved RM yields 
a viable measurement of black hole mass.
We also emphasize that,  as seen from the top right panel of Figure \ref{fig_factor}, the peak of the 
transfer function needs not coincide with the BLR size.

With a non-linearity parameter of $\gamma=0.3\pm0.04$, the BLR deviates 
slightly deviating linear response. Indeed, in support of this result, the 
excess variance of H$\beta$, $F_{\rm var}=0.169$, exceeds that of the
5100 {\AA} continuum flux, $F_{\rm var}=0.120$  (see Table 8 of \citealt{Bentz2009b}).  
Here, the excess variance is defined as 
$F_{\rm var}=\sqrt{\sigma_{F}^2-\delta_{F}^2}/\langle F\rangle$,
where $\sigma_{F}^2$ is the variance of the observed flux, $\delta_{F}^2$ is 
the mean square uncertainty, and $\langle F\rangle$ is the mean flux.
$F_{\rm var}$ represents the amplitude of flux variance of the light curve.

%
\begin{figure}[t!]
\centering
\includegraphics[angle=-90.0, width=0.45\textwidth]{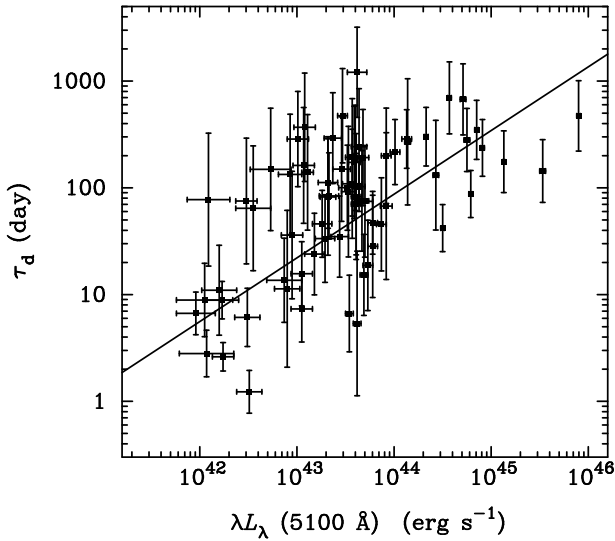}
\caption{Variation of the damping timescale $\tau_{\rm d}$ with optical 
(5100 \AA) AGN continuum luminosity.  Solid line represents the best linear fit.}
\label{fig_tau}
\end{figure}
%

%
\subsection{Comparisons with Other Independent Measurements}
{\it NGC 3227}. NGC 3227 is a local Seyfert 1 galaxy, sufficiently close
that its nuclear region is spatially resolved and dynamical modeling has been successfully 
applied to determine its central black hole mass. Using VLT SINFONI adaptive optics 
integral-field spectroscopy with a resolution of $\sim7$ pc, \cite{Davies2006} 
derived a black hole mass in the range $(0.7-2)\times10^7\,M_\odot$ from stellar 
dynamical modeling. Meanwhile, \cite{Hicks2008} analyzed the kinematics of the molecular hydrogen 
gas in the central region of NGC 3227 by modeling the gas velocity field
as a flat circular disk and reported a black hole mass of $2.0^{+1.0}_{-0.4}\times10^7\,M_\odot$. 

In our modeling of the RM data from the campaign at MDM observatory (\citealt{Denney2010}), 
the best value for the BLR size is $R_{\rm d}=5.1\pm0.7$ light days and the inclination 
angle is $\theta_{\rm inc}=60^\circ\pm22^\circ$. The  opening angle is 
$\theta_{\rm opn}=48^\circ\pm24^\circ$, but it is not well recovered because the distribution 
is nearly uniform between $0^\circ$ and $90^\circ$. This geometry roughly leads to a virial 
factor $f_{\rm BLR}\approx1-3$. As a result, the black hole mass for NGC 3227
lies in the range $(1.3-6.3)\times10^6\,M_\odot$ using the H$\beta$ line dispersion and  
$(2.2-10.6)\times10^6\,M_\odot$ using FWHM. These results are marginally consistent 
with previous dynamical mass measurements. Moreover, the H$\beta$ line in the rms spectra shows 
a prominent double-peak profile (see Figure 5 of \citealt{Denney2010}), strongly indicating a 
large inclination angle.

{\it NGC 4151}. NGC 4151 is another local Seyfert 1 galaxy with a measured central black hole mass.  
\cite{Hicks2008} used gas kinematics to derive a mass of 
$3.0^{+0.75}_{-2.2}\times10^7\,M_\odot$. The stellar dynamics
analysis of \cite{Onken2007} gave an upper limit to the black 
hole mass of $<5\times10^7\,M_\odot$. 

The latest monitoring campaign on NGC 4151 was carried out in 2006 using the 
1.3-m telescope
at MDM observatory (\citealt{Bentz2006}; see also \citealt{Maoz1991, Kaspi1996}).
Our modeling gives $R_{\rm d}=7.9\pm0.9$, $\theta_{\rm inc}=53^\circ\pm18^\circ$,
and $\theta_{\rm opn}=57^\circ\pm21^\circ$. For a virial factor of $f_{\rm BLR}\approx 1-3$, 
the black hole mass lies in a range of $(1.0-3.7)\times10^7\,M_\odot$ using $\sigma_{\rm line}$ and 
$(0.7-2.8)\times10^7\,M_\odot$ using FWHM. It is interesting to mention that
the stellar velocity dispersion of NGC 4151 is $\sigma_\star=98~{\rm km~s^{-1}}$
(\citealt{Onken2007}), which yields a black hole mass of $1.4\times10^7\,M_\odot$
according to the $M_\bullet-\sigma_\star$ calibration of \cite{Kormendy2013}.

In summary, the use of H$\beta$ light curves alone does not yield an accurate 
determination of the virial factor.  Nevertheless, tentative constraints on 
the virial factor yield black hole mass rough estimates compatible with 
independent mass estimates based on spatially resolved stellar and gas 
dynamical analysis.  Future incorporation of line profiles
will help to determine the virial factors more precisely.

%
\begin{figure}[t!]
\centering
\includegraphics[angle=-90.0, width=0.45\textwidth]{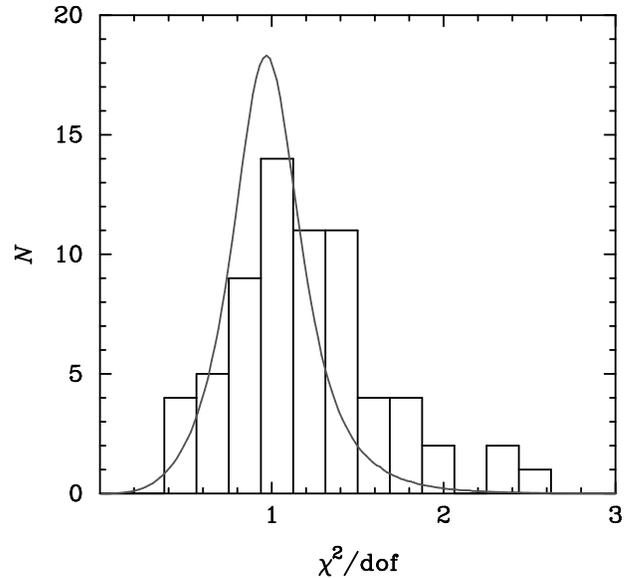}
\caption{Distribution of reduced $\chi^2$ for the emission-line fits. 
Gray solid line represents the expected $\chi^2$ distribution.}
\label{fig_chi2}
\end{figure}
\begin{figure}[t!]
\centering
\includegraphics[angle=-90.0, width=0.45\textwidth]{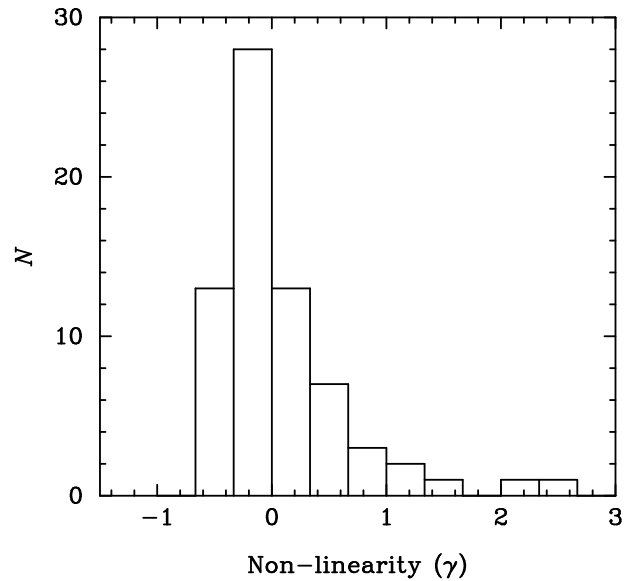}
\caption{Distribution of the non-linearity parameter $\gamma$ for the emission-line response to the continuum.}
\label{fig_gam}
\end{figure}
%

\section{Results}

\subsection{Overview}

In Figure \ref{fig_cases}, we also present illustration cases for NGC 5548 and 
NGC 3516.  The fit to the H$\beta$ light curve of NGC 5548 is remarkably good after detrending, 
as shown by the gray line. In particular, the small bump near 
JD54265 is well reproduced.  The corresponding transfer functions for Arp 151, 
NGC 5548, and NGC 3516 have quite diverse shapes, ranging from
a single peak, to roughly a top hat, to double peaks, respectively. 
For non-unimodal transfer functions like those of NGC 5548 and NGC 3516, the 
peak value  of the transfer function cannot be used to represent the BLR size.

The recovered parameters for the BLRs of all the objects are summarized in 
Table \ref{tab_par}.  We tabulate the damping timescale $\tau_{\rm d}$ but 
omit the parameter $\sigma_{\rm d}$ 
for the damped random walk model because we pay no attention to the absolute 
units of the light curves. Of the seven free parameters for the BLR model, 
$\beta$ and $F$, which jointly determine the width of the BLR, are almost 
uniformly distributed over the specified ranges. This implies that BLR widths 
$\sigma_R\approx 0.3R_{\rm BLR}$ can generally reproduce the observed H$\beta$ 
light curves. The virial factors are calculated according to Equation 
(\ref{eqn_factor}) for $\theta_{\rm opn}<40^\circ$. The uncertainties are not 
given owing to the approximation underlying Equation (\ref{eqn_factor}). 
Overall, the virial factors range from $\sim1$ up to $\sim10$. Future 
velocity-resolved RM will better constrain the virial factors.

In Appendix B, we provide a supplementary online figure to summarize the 
reconstructions of the continua and emission lines, the derived transfer 
functions, and the distributions of BLR parameters for all the objects. We 
present the detailed results in following sections.

\subsection{Damped Random Walk Model for the Continuum}

Figure \ref{fig_tau} shows the dependence of the rest-frame damping timescale 
for the damped random walk model for AGN variability on the luminosity at 
5100 \AA. As expected, the timescale $\tau_{\rm d}$ increases with luminosity 
(\citealt{Kelly2009, MacLeod2010}). A linear regression using the fitting 
algorithm FITEXY (\citealt{Press1992}, including a term for the intrinsic 
scatter; see Section \ref{sec_rl} below for details),  yields
\begin{equation}
 \log \left(\frac{\tau_{\rm d}}{\rm day}\right) = (1.94\pm0.07) + (0.60\pm0.06) \log 
  \left[\frac{\lambda L_{\lambda}~(5100\,\text{\AA})}{10^{44}~\rm erg~s^{-1}}\right].
\end{equation}
A previous study by \cite{Kelly2009} found a similar relation using a sample 
composed of $\sim$55 high-redshift AGNs from the MACHO survey and 45 nearby 
AGNs. Interestingly, \cite{Kelly2009} claimed that this relationship is 
consistent with that for the orbital or thermal timescale of the accretion 
disks, and thus a constraint on the viscosity parameter can be obtained. Given 
the feasibility of obtaining robust mass measurements from the present RM 
sample, a comprehensive investigation of the correlation between the 
parameters of damped random walk model with black hole mass will provide 
insights into the physics of accretion disks and the origins of their 
variability (\citealt{Dexter2011}). 

%
%
\begin{figure}[t!]
\centering
\includegraphics[angle=-90.0, width=0.45\textwidth]{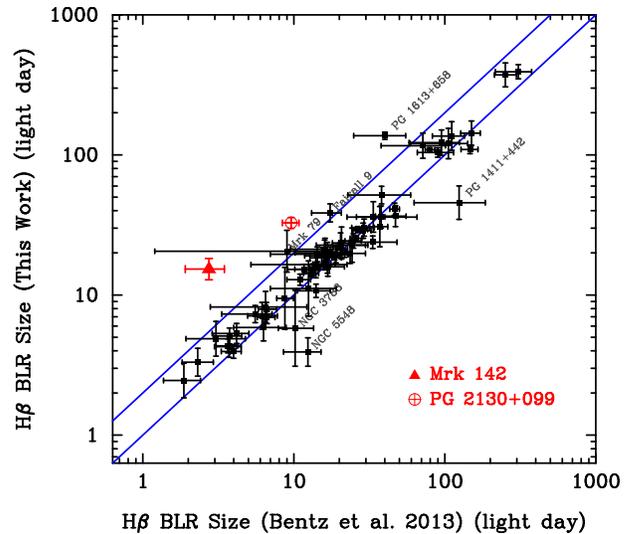}
\caption{Comparison of the H$\beta$ BLR size derived from this work with that from 
\cite{Bentz2013}. The two solid lines represent $y=x$ and $y=2x$, respectively.}
\label{fig_blrsize}
\end{figure}
%

%
%
\subsection{Response of Emission Lines}

Figure \ref{fig_chi2} presents the distribution of the reduced $\chi^2$ for
the emission-line fits using the best recovered parameter set. 
The expected distribution of $\chi^2$/dof is also superposed for comparison. 
There is a moderate discrepancy for $\chi^2$/dof larger that 
$\sim1.5$; this is probably due to systematic errors 
(e.g., calibration errors) that may not included in the 
reported errors of the original data (e.g., NGC 4051 and PG 0052+251).
Among the set of 70 light curves analyzed, the fits are especially poor 
($\chi^2/{\rm dof}>3$) for PG 1613+658 and Mrk 79 (during JD48149--JD48345 and 
JD48905--JD49135).
This is not surprising since the H$\beta$ light curves of these three cases
are sparsely sampled and suffer the most acute systematic errors. Nevertheless, the 
calculated $\chi^2$ distribution is generally consistent with the expectation, 
indicating that our modeling reasonably reproduces the emission-line data.

Figure \ref{fig_gam} plots the distribution of the non-linearity parameter $\gamma$. 
Although the mean value of $\gamma$ is close to 0, several objects exhibit prominent 
deviations. Indeed, it is common for the variability amplitude of the 
emission lines to exceed those of the continuum (\citealt{Meusinger2011} and references therein). 
This can be simply verified by inspection of the ``excess variance'' values,
which are usually provided in the literature. The non-linear response of
H$\beta$ might be ascribed to the following reasons:
(1) The unobservable ionizing UV continuum is not linearly
correlated with optical continuum (i.e., the shape of the spectral energy distribution
is changing along with the continuum variation).  This is plausible in light of the fact that,
according to the standard accretion disk theory (\citealt{Shakura1973}), the 
UV and optical emission mainly comes from different regions.  (2) The portion 
of the BLR emitting H$\beta$ may be (partially) optically thick so that the 
line intensity depends on the ionization parameter (e.g., \citealt{Netzer1990}).
However, in such a situation, H$\beta$ emission may be anisotropic. We do not 
consider this effect in the present modeling. As mentioned above, RM 
observations of NGC 5548 tend to support the first explanation 
(\citealt{Peterson2002, Bentz2007}), although the second one cannot be excluded.  

%
\begin{figure}[t!]
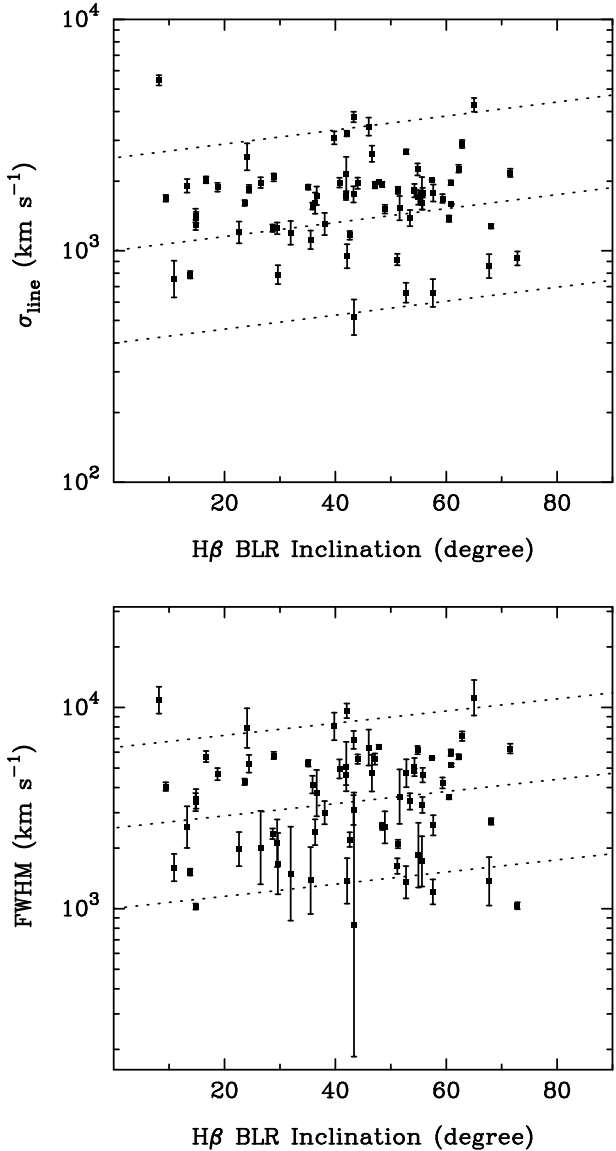

\centering
\includegraphics[angle=-90.0, width=0.45\textwidth]{fig_inclination_sig.ps}\\
\vspace*{0.5cm}
\includegraphics[angle=-90.0, width=0.45\textwidth]{fig_inclination_fwhm.ps}
\caption{Dependence of H$\beta$ (top) line dispersion and (bottom) FWHM 
measured from rms spectra with inclination angle.  The uncertainties for the 
inclination angles are typically $20^\circ-30^\circ$ and are not shown for 
clarity.  To guide the eye, the dotted lines plot the relation $V_{\rm obs} = 
V_{\rm Kep}(\sin^2\theta_{\rm opn} + \sin^2\theta_{\rm inc})^{1/2}$
with $\theta_{\rm opn}=40^\circ$ and $V_{\rm Kep}=10^{2.8}$, $10^{3.2}$, and 
$10^{3.6}~\rm km~s^{-1}$ in the top panel and $V_{\rm Kep}=10^{3.2}$, 
$10^{3.6}$, and $10^{4.2}~\rm km~s^{-1}$ in the bottom panel.}
\label{fig_inc}
\end{figure}
%

%
\subsection{Broad-line Region Geometry}

Figure \ref{fig_blrsize} compares the H$\beta$ BLR sizes derived from the present 
approach with those from the traditional analysis via cross-correlation functions 
(CCFs) between the continuum and the emission lines. We use the homogeneous 
compilation of CCF-based results from \cite{Bentz2013}. Except for three 
objects PG 1411+442, NGC 5548 (JD54180--JD54332), and NGC 3783, the BLRs sizes 
derived using our approach are all systematically larger.  Quantitatively, our 
derived BLR sizes are on average larger than CCF-based sizes by $\sim$20\%. 
This confirms previous suspicions that CCF analysis might 
underestimate the sizes of BLRs (\citealt{Netzer1990, Maoz1991, Welsh1999}).

As discussed by \cite{Maoz1991}, CCF analysis tends to weigh the inner parts 
of the BLR more than the outer parts%
\footnote{However, this does not at all mean that the current black hole mass 
measurements based on the $R_{\rm BLR}-L$ relationship are problematic. Note 
that the virial factors are calibrated using the lag determinations from 
CCF analysis.}.
The derived time lags will be considerably biased when the transfer functions 
are broad and multimodal. It is straightforward to derive the relation between 
the CCF and the transfer function:
\begin{eqnarray}
{\rm CCF}(\tau)&=&\int f_c(t)f_l(t+\tau)\rd t\nonumber\\
&=&A\int f_c(t)\int f_c^{1+\gamma}(t+\tau-\tau')\Psi(\tau')\rd\tau'\rd t\nonumber\\
&=&A\int \Psi(\tau')\int f_c(t)f_c^{1+\gamma}(t+\tau-\tau')\rd\tau'\rd t.
\end{eqnarray}
Regardless of the non-linearity parameter $\gamma$ for now, we obtain a 
concise form (e.g., \citealt{Welsh1999}),
\begin{equation}
{\rm CCF}(\tau)= \int \Psi(\tau') {\rm ACF}_c(\tau-\tau')\rd\tau',
\end{equation}
where ${\rm ACF}_c$ refers to the autocorrelation function of the continuum 
\begin{equation}
{\rm ACF}_c(\tau-\tau')=\int f_c(t) f_c(t+\tau-\tau')\rd t.
\end{equation}
Consider an idealized case: ${\rm ACF}_c$ is a delta function so that the CCF 
is exactly identical to the transfer function. Consequently, multiple peaks in 
the transfer function (e.g., of an inclined disk-like BLR) lead to ambiguity 
in determining the peak values of the CCF. To be more specific, a thick, 
inclined disk has two peaks in the transfer function, and the stronger one is 
located closer in than the weaker one (e.g., see Figure \ref{fig_cases}). 
Hence, using the peak value of the CCF to determine the BLR size in such a 
circumstance biases the estimate toward the inner radius. A realistic 
situation will be even far more complicated once we account for the
finite-duration, irregular sampling and observational noise.

Regarding the three cases whose size estimates may have been biased too low, 
it is too early to draw firm conclusions for PG 1411+442 and NGC 3783 because 
their light curves are sparsely sampled and noisy. However, for NGC 5548 
(JD54180--JD54332), inspection of the CCF given by \cite{Denney2010} shows 
that it is quite broad, despite the high quality of the data set (see their 
Figure 3). This leads to a correspondingly broad transfer function
(Figure \ref{fig_cases}), such that the CCF measurement can easily be skewed 
to incorrect values. Interestingly, the centroid value (12.4~days)
of the CCF is twice as larger as the peak value (6.1 days).

In Figure \ref{fig_inc}, we show the inclination dependence of the H$\beta$ 
line widths measured from rms spectra as line dispersion and FWHM.  The error 
bars for the inclination angles are typically $20^\circ-30^\circ$ and are not 
plotted for clarity.  To guide the eye, we superpose the relation described by 
Equation (\ref{eqn_width}) with $\theta_{\rm opn}=40^\circ$ and 
$V_{\rm Kep}=10^{2.8}$, $10^{3.2}$, and $10^{3.6}~\rm km~s^{-1}$ in the top 
panel and $V_{\rm Kep}=10^{3.2}$, $10^{3.6}$, and $10^{4.2}~\rm km~s^{-1}$ in 
the bottom panel. The relationship between line width and inclination angle 
has quite a large scatter because of the wide range of opening angles and 
$V_{\rm Kep}$, which depends on the central black hole mass and the BLR size.  
Nevertheless,  it does seem that line width does increase slightly with 
inclination angle, in qualitative agreement with Equation (\ref{eqn_width}), 
suggesting that the derived inclination angles are probably meaningful.

\subsection{Notes on Two Individual Objects}

We here demonstrate the ability of the present Bayesian approach to properly
derive the BLR sizes for Mrk 142 and PG 2130+099, both of which were 
previously considered to have unreliable lag determinations from CCF analysis 
because they were significant outliers in the $R_{\rm BLR}-L$ relation 
(\citealt{Bentz2009b, Grier2012}). 

{\em Mrk 142}. This object was monitored by the LAMP project with a duration 
of $\sim100$ days (\citealt{Bentz2009b, Walsh2009}). The data for the H$\beta$ 
line were suspected to suffer from large systematic errors, probably due to 
Fe{\footnotesize II} blending with [O{\footnotesize III}], which was used to 
calibrate the H$\beta$ fluxes (\citealt{Bentz2009b}). The CCF analysis by 
\cite{Bentz2009b} yielded an H$\beta$ time lag of 2.87 days with respect to 
the $V$-band continuum, making Mrk 142 deviate strongly from the 
$R_{\rm BLR}-L$ relationship (\citealt{Bentz2013}).  However, our modeling 
gives a lag of $15.3\pm2.7$ days, a factor of six larger than reported by
\cite{Bentz2009b}.  The BLR size for Mrk 142 derived here
is remarkably consistent with the $R_{\rm BLR}-L$ relationship.

{\em PG 2130+099}. This object was first monitored by \cite{Kaspi2000}, who
measured an H$\beta$ time lag of $188$ days, with a large uncertainty. 
Inspection of the H$\beta$ light curve of Kaspi et al. (see their Figure 3)
shows that a seasonal gap of a period of $\sim200$ days may dominate the CCF 
and alias the lag determination.  Later studies by \cite{Grier2012} with 
fairly well-sampled data determined a lag of 12.8 days, which falls well 
below the $R_{\rm BLR}-L$ relation (see also discussion in
\citealt{Bentz2013}). However, a subsequent comprehensive investigation of the 
same data set by \cite{Grier2013b} using the maximum entropy technique found 
that the transfer function (or delay map) contains two peaks. The stronger one 
is located around 12.5 days and the weaker one at 31 days; therefore, the 
possibility of a lag of $\sim31$ days cannot be excluded.

In our analysis, a thin disk model can reproduce the H$\beta$ line fluxes with 
$\chi^2$/dof=1.19.  The inclination and opening angle are fairly well 
recovered: $\theta_{\rm inc}=51^\circ\pm6^\circ$
and $\theta_{\rm opn}<18^\circ$. There are two peaks in the distribution of 
BLR size, located at $\sim$21 and 33 days; the latter peak has significantly 
higher probability. On the other hand, it is also possible that PG 2130+099 
undergoes secular variations.  After applying our detrending procedure, we 
find a BLR size of $21.0\pm4.0$ with a $\chi^2$/dof=1.31 (see online Figure~\ref{fig_online}).
We adopt a BLR size of $32.7\pm3.0$ days, which places PG 2130+099 much
closer to the $R_{\rm BLR}-L$ relationship. 

%
\begin{figure}[t!]
\centering
\includegraphics[angle=-90.0, width=0.45\textwidth]{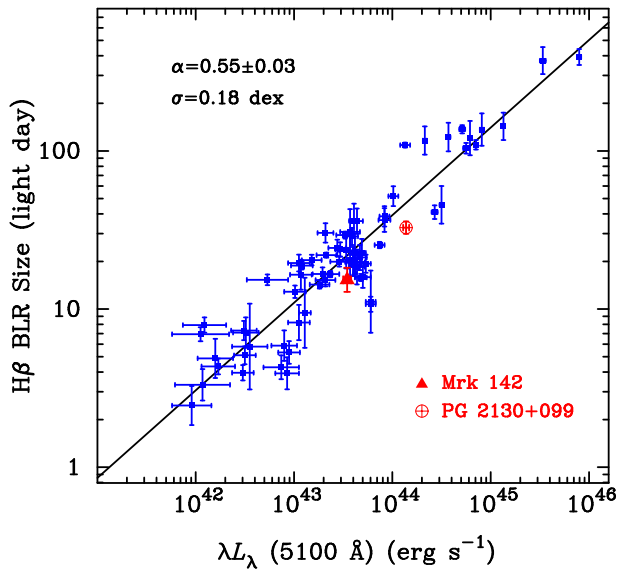}
\caption{Relation between H$\beta$ BLR size and 5100 \AA~luminosity. 
Solid line is the best linear fit.}
\label{fig_rl}
\end{figure}
%

%
%
\section{Implications for the $R_{\rm BLR}-L$ Relation}
\label{sec_rl}
The relationship between H$\beta$ BLR size and 5100 \AA~luminosity is plotted in 
Figure \ref{fig_rl}.  As in \cite{Kaspi2005}, we use the fitting routine 
FITEXY (\citealt{Press1992}) to include an intrinsic scatter, following the 
prescription of \cite{Tremaine2002}.  Specifically, we add in quadrature the 
uncertainty of $R_{\rm BLR}$ until we obtain a reduced $\chi^2$ of unity.  The 
best fit gives 
\begin{equation}
\log \left(\frac{R_{\rm BLR}}{\rm ld}\right) = K + 
   \alpha \log\left[\frac{\lambda L_\lambda (5100\,\text{\AA})}{10^{44}\rm erg~s^{-1}}\right],
\end{equation}
with $K=1.59\pm 0.02$ and $\alpha=0.55\pm 0.03$. The scatter of the fit
is 0.18 dex, comparable to the 0.19 dex reported by \cite{Bentz2013}.  If our 
current analysis correctly models the geometry of the BLR, our derived 
$R_{\rm BLR}-L$ relation should account for the effect of inclination. That our 
inferred intrinsic scatter for the $R_{\rm BLR}-L$ relation is comparable to 
that of previous studies that did not include the inclination effect seems to 
imply that the scatter is dominated by other factors.

Intensive studies on NGC 5548 suggest that the slope of the $R_{\rm BLR}-L$ 
relation for individual objects may depart significant from the average slope. 
Using the 13-year optical and UV monitoring data of NGC 5548, 
\cite{Peterson2002} found that the H$\beta$ lag correlates with UV flux as 
$\tau\propto F_{\rm UV}^{0.53}$, but, surprisingly, that it correlates almost 
linearly with the optical flux, $\tau\propto F_{\rm opt}^{0.93}$.  If this 
behavior also holds for other RM objects, it should contribute a major source 
of scatter to the $R_{\rm BLR}-L$ relation.  On the other hand, from a 
theoretical point of view, the shape of the emergent spectrum from the 
accretion disk depends on black hole mass, accretion rate, and as well as 
black hole spin.  As the RM sample spans several orders of magnitude in 
luminosity, the correlation between the optical and ionizing UV luminosity 
should be quite diverse and hence must be a key source of scatter for the 
$R_{\rm BLR}-L$ relation.  Other factors, such as Eddington ratio, may also 
contribute.

The best-fit slope of $\alpha=0.55\pm0.03$ is close to the value of 
$\alpha=0.53\pm0.03$ reported by \cite{Bentz2013}, but marginally different
from the expectation of 0.5 based on simple photoionization arguments. This is 
not surprising in light of the use of the optical continuum instead of 
the ionizing UV continuum and the non-linear response of the line emission,
which implies that the shape of the incident optical-UV continuum might 
be changing with reverberation variations or, alternatively, that the BLR is 
optically thick to H$\beta$. Both of these two possibilities violate the 
simplistic assumptions of the photoionization argument. As mentioned above, 
long-term monitoring of NGC 5548 provides evidence that the shape of the 
continuum does change (\citealt{Peterson2002}; see also \citealt{Bentz2007}).

\section{Discussion}
We have demonstrated the fidelity of our approach for diagnosing the structure 
of the BLR, in particular in its ability to reproduce the light curves of 
emission line fluxes. Compared with the original framework developed by 
\cite{Pancoast2011}, we explicitly include the non-linear 
response of the line emission to the continuum and detrending of the light 
curves to remove long-term secular variability sometimes seen in both the 
emission lines and continuum. The transfer function is self-consistently
calculated in the present approach, obviating the need assume it, as in 
\cite{Zu2011}. However, we do need to specify a general BLR model in advance
for the implementation of the Bayesian analysis.  Thus, we cannot claim that 
the BLR model is unique.  In the future, we can circumvent this limitation by 
model selection when the observations are of sufficient quality to justify
construction of more refined BLR models. 

Previous studies that have succeeded in recovering velocity-delay maps suggest 
that multiple components coexist in the BLR, including disks, winds, and 
inflows/outflows (e.g., \citealt{Bentz2010, Grier2013b}; see also 
\citealt{Peterson2013} for a review).  This does not invalidate the present 
modeling, which makes use of only the H$\beta$ light curves. Only with 
velocity-resolved mapping data can these different kinematics be 
distinguished. Without any additional information on the preferred geometry of 
the BLR, we proceed with the assumption that the BLR has a flexible disk 
geometry. 

Future inclusion of emission line profiles will definitely provide more 
stringent constraints on the BLR geometry in general, and on the inclination 
angle in particular (see the works of \citealt{Brewer2011} and 
\citealt{Pancoast2012}). It will then be instructive to try to compare the 
inclination of the BLR with indicators of the orientation of the accretion 
disk (e.g., \citealt{Nishiura1998, Wu2001, Jarvis2006, Runnoe2013}).  This 
will allow us test models for the formation and evolution of the BLR (see 
\citealt{Wang2011, Wang2012} and references therein). The most promising 
indicator of the orientation of the accretion disk is the relativistically 
broadened Fe K$\alpha$ emission line, which is believed to originate from the 
inner region of the accretion disk and hence should trace the overall 
inclination of the disk (e.g., \citealt{Patrick2012, Walton2013} and 
references therein). The heuristic work by \cite{Nishiura1998} correlated the 
disk inclination, derived from early rudimentary modeling of broad Fe 
K$\alpha$, with emission-line widths and concluded that the BLR may arise from 
the outer parts of a warped accretion disk.

The present approach might fail to reproduce the correct BLR size in
cases where the light curve is poorly sampled or suffers from severe seasonal 
aliasing (e.g., 3C 120 in \citealt{Peterson1998}).  However, these cases can 
be easily singled out through simple visual inspection.  Supplying Equation 
(\ref{eqn_prob}) with more sophisticated priors or using sections of the light 
curve with relatively better sampling rates helps to eliminate seasonal 
aliasing.  In addition, as suggested by \cite{Zu2011}, using multiple emission 
lines can further alleviate sampling problems and seasonal aliasing. For this 
purpose, one needs to simultaneously fit the continuum and multiple emission 
lines (\citealt{Zu2011}), and one needs to treat the entire continuum as 
unknown but to be inferred from the data (\citealt{Pancoast2011}). This will 
be computationally challenging and is unfeasible for analyzing a large sample. 
We are currently developing a parallelization of our present approach 
using a supercomputer cluster to address these issues.

It is worth pointing out that the size and structure of
BLRs potentially vary with emission lines. RM studies of various emission lines
see evidence for stratification of the BLR; namely, low-ionization lines 
(such as H$\beta$ and C~{\footnotesize III}) respond with longer time delays 
than higher ionization lines (such as He~{\footnotesize  II} and 
C~{\footnotesize IV}; e.g., \citealt{Clavel1991, Dietrich1995, Kollatschny2003}).
This indicates that BLRs are obviously much more complicated than 
our simple modeling. An extension of the present approach by 
self-consistently including multiple emission lines will shed light 
on the global structures of BLRs.

Apart from the above points, there are several other aspects for future 
improvement. 

\begin{enumerate}
 \item For simplicity, we assume that the continuum fluxes are isotropic and 
decline as the inverse square of the distance of clouds from the central 
ionizing source. More sophisticated cases are beyond the goal of this work, 
but are worth studying in the future with high-fidelity data.

 \item Incorporate realistic and physical ionization calculations
(e.g., using the well-developed package CLOUDY; \citealt{Ferland1998}) to take 
into account the effects of optical depth and probe the physical environment 
of the BLR (\citealt{Chiang1996, Ferland2009}; see also \citealt{Horne2003}).
 
 \item Include the influence of radiation pressure on the kinematics of the 
BLR clouds; radiation pressure is believed to lead to underestimates of the 
black hole mass for objects with high Eddington ratios
 (\citealt{Marconi2008, Netzer2009}).
 
 \item Construct physical models for inflows and outflows, plausible 
components inferred in a number of previous studies (e.g., 
\citealt{Kollatschny1996, Gaskell2013}). 

\end{enumerate}

%
%
\section{Conclusions}
Motivated by recent advances in BLR modeling and statistical description of 
AGN variability, we carry out a systematic study of BLR structure using all 
the RM data with H$\beta$ monitoring available in the literature. 
We improve on previous efforts by 
incorporating the non-linearity of the line response to the continuum and by
detrending the light curves for secular variability. Although a general disk 
geometry is assumed in this initial work, the flexibility of the Bayesian 
approach readily allows us in the future to account for more complicated 
BLR structures and physical processes. Our main results are as follows.

\begin{enumerate}
\item The damped random walk model can explain the variability of the 
optical continuum for all the RM objects, confirming the results of previous 
studies (e.g., \citealt{Kelly2009, Zu2013}). The advantage with RM sample is 
that the black hole mass can be simply estimated, thus permitting the physical 
origin of the variability to be investigated by linking it to the properties 
of the accretion disk (e.g., Eddington ratio, orbital motion, thermal 
processes).

\item The observed H$\beta$ light curves can be fairly well reproduced by 
a general geometry for the BLR that accounts for disks, rings, shells, and 
spheres.  This indicates that the structure of the BLR 
for H$\beta$ emission line is mainly disk-like.

\item The H$\beta$ BLR sizes determined here through our Bayesian method are 
systematically larger by $\sim20$\% compared to those derived from the
traditional cross-correlation analysis. This discrepancy plausibly arises 
from the fact that the cross-correlation method biases is biased toward the 
inner parts of the BLR (\citealt{Netzer1990, Maoz1991, Welsh1999}).

\item We redetermine the $R_{\rm BLR}-L$ relationship and find a slope of 
$0.55\pm0.03$ and an intrinsic scatter of 0.18 dex. Since the derived BLR 
sizes have already taken into account the inclination effect, the remaining 
scatter must arise from other factors, including variation in the shape 
of the continuum and diverse Eddington ratios.

\item In present framework, we find that the non-linear response of the 
H$\beta$ line emission to the continuum is required to better reproduce the 
observed line fluxes (see Figure~\ref{fig_gam}).  This seems reasonable, 
in light of the fact that the variation amplitude of the line emission 
sometimes exceeds that of the continuum. The non-linearity may be ascribed to 
the non-linear correlation between the optical and ionizing UV continuum 
(\citealt{Peterson2002}) or to the existence of (partially) optically thick 
BLR clouds.

\item We demonstrate the capability of the present approach to recover 
appropriate BLR sizes when the traditional cross-correlation analysis 
fails.  The newly obtained BLR sizes for Mrk 142 and PG 2130+099, previously 
reported to be major outliers of the $R_{\rm BLR}-L$ relationship 
(\citealt{Bentz2013}), shows remarkable consistency with the 
anticipated values (see Figure \ref{fig_rl}).
\end{enumerate}

\acknowledgements{We thank the members
of IHEP AGN group for discussions. This research is supported by NSFC-11133006, 11173023, 11233003, and
11303026, a 973 project (2009CB824800), and the China-Israel NSFC-ISF 11361140347.  The work of L.C.H. is 
supported by the Kavli Foundation, Peking University, and the Carnegie 
Institution for Science.}

\begin{figure*}[th!]
\includegraphics[angle=-90.0, width=0.95\textwidth]{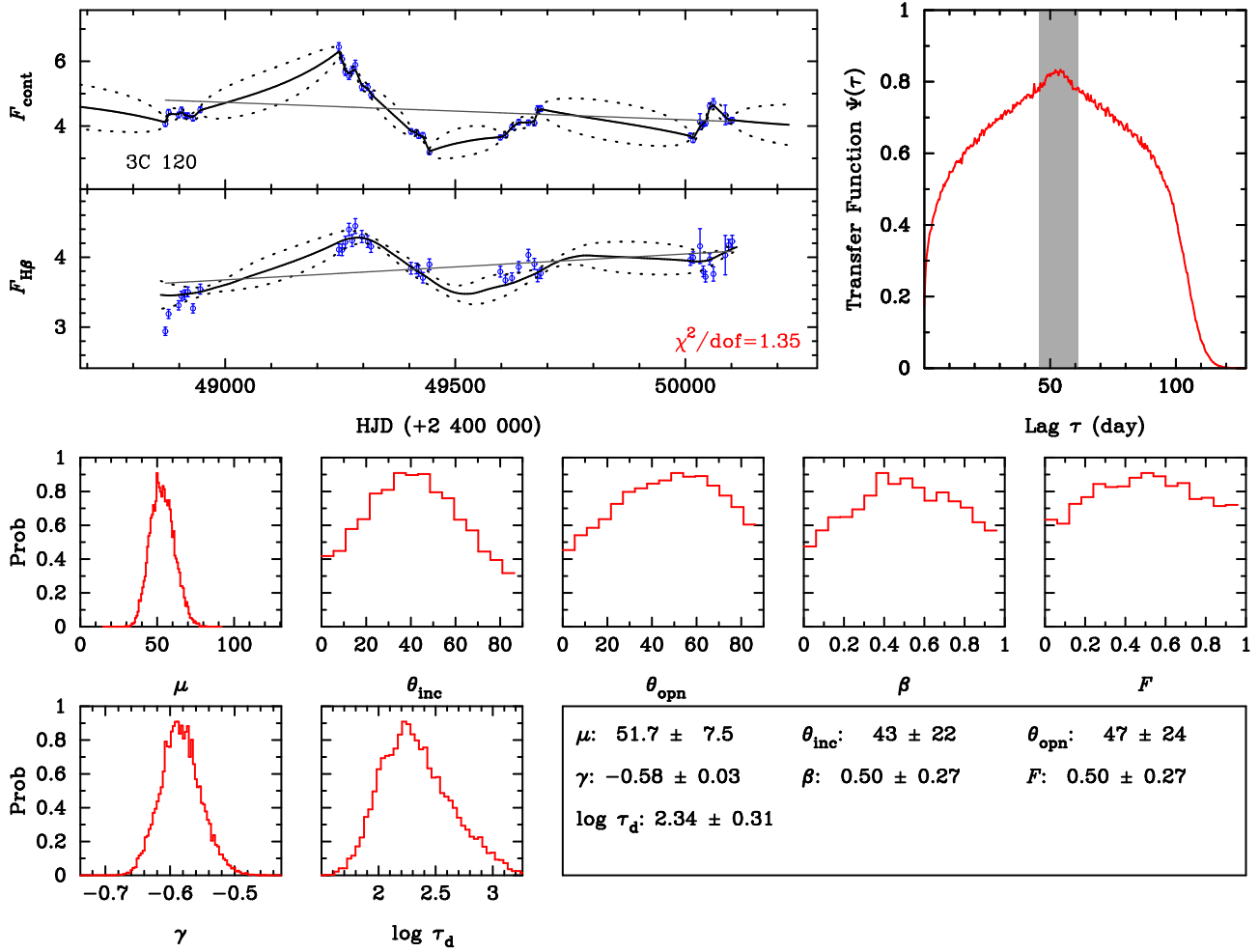}
\caption{Continuum reconstructions and BLR modelings. 
({\em Top left}) Data points with error bars are the observed light curves. 
Thick solid lines show the reconstructed continuum at 5100 {\AA} and the recovered 
H$\beta$ emission; dashed lines represent the uncertainties. 
If a detrending is initiated, the gray lines show the fits for the detrending
applied to remove the secular variations. ({\em Top right}) Transfer function $\Psi(\tau)$ 
(red curve, in arbitrary units) derived from the best-fit BLR model. The vertical shaded area represents 
the recovered BLR size and its uncertainty in observer's frame.
({\em Bottom}) Distributions for the parameters of mean radius $\mu$ (light day),
inclination angle $\theta_{\rm inc}$ (degree), opening angle $\theta_{\rm opn}$ (degree),
$\beta$, $F$, non-linearity $\gamma$, and typical damping timescale $\tau_{\rm d}$ (day).
Their statistical values are summarized in bottom right panel. Note that here
$\mu$ and $\tau_{\rm d}$ are given in the rest-frame.}
\label{fig_online}
(The complete figure set (71 images) and color version are available in the online journal.)
\end{figure*}

\appendix
\section{Equations for the Continuum Reconstruction}
\label{app_eqn}
We derive the equations for the continuum reconstruction described in Section \ref{sec_continuum}. 
Assuming that both the signal $\mathbi{s}$ and the noise $\mathbi{n}$ have a Gaussian distribution with 
covariance matrices $\mathbi{S}=\langle \mathbi{ss}^T\rangle$ and $\mathbi{N}=\langle \mathbi{nn}^T\rangle$, 
respectively,  their probability distributions are
\begin{equation}
P(\mathbi{s})\propto|\mathbi{S}|^{-1/2}\exp\left(-\frac{\mathbi{s}^T\mathbi{S}^{-1}\mathbi{s}}{2}\right),
\end{equation}
and 
\begin{equation}
P(\mathbi{n})\propto|\mathbi{N}|^{-1/2}\exp\left(-\frac{\mathbi{n}^T\mathbi{N}^{-1}\mathbi{n}}{2}\right),
\end{equation}
where the angle bracket denotes statistical ensemble average and the superscript ``$T$'' denotes the 
transposition of vectors or matrices. We further assume that $\mathbi{s}$ and $\mathbi{n}$ 
are uncorrelated, so that the joint probability $P(\mathbi{s, n})=P(\mathbi{s})P(\mathbi{n})$. 
Now we calculate the probability for the observations $\mathbi{y=s+n+E}q$ as (\citealt{Rybicki1992})
\begin{eqnarray}
P(\mathbi{y}|q, \sigma_{\rm d}, \tau_{\rm d}) 
&=&\int P(\mathbi{s, n})\delta\left[\mathbi{y-(s+n+E}q)\right]\rd^m \mathbi{s} \rd^m \mathbi{n}\nonumber\\
&=&\int P(\mathbi{s})P(\mathbi{n=y-E}q-\mathbi{s})\rd^m \mathbi{s}\nonumber\\
&\propto& \exp\left[-\frac{1}{2}(\mathbi{y-E}q)^T(\mathbi{S+N})^{-1}(\mathbi{y-E}q)\right]
         \int \exp\left[-\frac{1}{2}(\mathbi{s-\hat s)}^T(\mathbi{S}^{-1}+\mathbi{N}^{-1})(\mathbi{s-\hat s})\right]
         \rd^m \mathbi{s},
\end{eqnarray}
where $\mathbi{\hat s}=\mathbi{S(S+N)}^{-1}\mathbi{(y-E}q)$ and the equalities 
$(\mathbi{S}^{-1}+\mathbi{N}^{-1})^{-1}=\mathbi{S}(\mathbi{S+N})^{-1}\mathbi{N}=\mathbi{N}(\mathbi{S+N})^{-1}\mathbi{S}$ 
are used. Substituting the variable of integration with $(\mathbi{s-\hat s)}$ and manipulating a simple 
normalization yields
\begin{equation}
P(\mathbi{y}|q, \sigma_{\rm d}, \tau_{\rm d}) 
= \frac{1}{\sqrt{(2\pi)^m|\mathbi{C}|}}\exp\left[-\frac{1}{2}(\mathbi{y-E}q)^T\mathbi{C}^{-1}(\mathbi{y-E}q)\right],
\end{equation}
where $\mathbi{C\equiv S+N}$. We marginalize the parameter $q$ by ``completing the square''
\begin{equation}
(\mathbi{y-E}q)^T\mathbi{C}^{-1}(\mathbi{y-E}q) = (\mathbi{y-E}\hat q)^T\mathbi{C}^{-1}(\mathbi{y-E}\hat q)
+(q-\hat q)\mathbi{E}^T\mathbi{C}^{-1}\mathbi{E}(q-\hat q),
\end{equation}
where 
\begin{equation}
\hat q = \frac{\mathbi{E}^T\mathbi{C}^{-1}\mathbi{y}}{\mathbi{E}^T\mathbi{C}^{-1}\mathbi{E}}.
\end{equation}
This formula is the generalized ``inverse-variance weighted mean'' that accounts for the 
correlation between data points in a way that a data value highly correlated with other data value 
gets a small weight (\citealt{Rybicki1992}). As a result, by assuming, as usual, that $P(q)$ is constant,
we arrive at
\begin{equation}
P(\mathbi{y}|\sigma_{\rm d}, \tau_{\rm d}) = \int P(\mathbi{y}|q, \sigma_{\rm d}, \tau_{\rm d}) P(q)\rd q 
= \frac{1}{\sqrt{(2\pi)^m|\mathbi{C}|}}\exp\left[-\frac{1}{2}(\mathbi{y-E}\hat q)^T\mathbi{C}^{-1}(\mathbi{y-E}\hat q)\right].
\end{equation}
From above manipulation, one readily finds the best estimate of $\mathbi{y}$:
\begin{equation}
\mathbi{\hat y} = \mathbi{\hat s+E}\hat q = \mathbi{SC}^{-1}\mathbi{(y-E}\hat q) + \mathbi{E}\hat q. 
\end{equation}
The mean square residual of the estimate is calculated by
\begin{eqnarray}
\left\langle\left[\mathbi{\hat y} - (\mathbi{s-E}q)\right]^2\right\rangle
&=& \left\langle\left[\mathbi{SC}^{-1}\mathbi{(y-E}\hat q) + \mathbi{E}\hat q - \mathbi{s}\right]
\left[\mathbi{SC}^{-1}\mathbi{(y-E}\hat q) + \mathbi{E}\hat q - \mathbi{s}\right]^T\right\rangle\nonumber\\
&=&\mathbi{S}-\mathbi{S}^T\mathbi{C}^{-1}\mathbi{S}
+\frac{(\mathbi{SC}^{-1}\mathbi{E-E})(\mathbi{SC}^{-1}\mathbi{E-E})^T}{\mathbi{E}^T\mathbi{C}^{-1}\mathbi{E}},
\end{eqnarray}
where the terms invoking $q$ are eliminated for ensemble averages. 

\section{Online Figures}
In the online version of Figure 10, we plot the reconstructions of the continuum light curves, the fits for the
H$\beta$ light curves, and the recovered parameter distributions of the BLR for all the objects in the sample, in the 
order as they appear in Table \ref{tab_obj}.

%

\LongTables
\begin{deluxetable}{lccccccc}
\tablecolumns{8}
\tablewidth{1.0\textwidth}
\tablecaption{Object Properties}
\tablehead{
  \colhead{Object} 
& \colhead{Redshift}    
& \colhead{Julian Dates}
& \colhead{$\log\lambda L_\lambda$ (5100 \AA)}
& \colhead{FWHM}
& \colhead{$\sigma_{\rm line}$}
& \colhead{Reference}\\
  \colhead{} 
& \colhead{}  
& \colhead{(+2 400 000)}
& \colhead{(erg s$^{-1}$)}
& \colhead{(km s$^{-1}$)}
& \colhead{(km s$^{-1}$)} 
& \colhead{}}
\startdata
3C 120             &   0.03301   & 48869-50101 &    $44.01\pm0.05~$     & $2205\pm 185\phn$   & $1166\pm 50\phn$      &  1 \\
\nodata            &   \nodata   & 55430-55569 &    $43.87\pm0.05~$     & $2539\pm 466\phn$   & $1514\pm 65\phn$      &  2 \\
3C 390.3           &   0.05610   & 49718-50012 &    $43.62\pm0.10~$     & $9630\pm 804\phn$   & $3211\pm 90\phn$      &  3 \\
\nodata            &   \nodata   & 53630-53713 &    $44.43\pm0.03~$     &$10872\pm1670$       & $5455\pm278$          &  4\\
Ark 120            &   0.03230   & 48149-48345 &    $43.92\pm0.06~$     & $5536\pm 297\phn$   & $1959\pm109$          &  1 \\
\nodata            &   \nodata   & 48870-49090 &    $43.57\pm0.10~$     & $5284\pm 203\phn$   & $1884\pm 48\phn$      &  1 \\
Arp 151            &   0.02109   & 54506-54607 &    $42.48\pm0.11~$     & $2357\pm 142\phn$   & $1252\pm 46\phn$      &  5, 6\\
Fairall 9          &   0.04702   & 50473-50665 &    $43.92\pm0.05~$     & $6901\pm 707\phn$   & $3787\pm197$          &  7, 8 \\ 
Mrk 79             &   0.02219   & 47838-48044 &    $43.57\pm0.07~$     & $5086\pm1436$       & $2137\pm375$          &  1 \\
\nodata            &   \nodata   & 48193-48393 &    $43.67\pm0.07~$     & $4219\pm 262\phn$   & $1683\pm 72\phn$      &  1 \\
\nodata            &   \nodata   & 48905-49135 &    $43.60\pm0.07~$     & $5251\pm 533\phn$   & $1854\pm 72\phn$      &  1 \\
Mrk 110            &   0.03529   & 48954-49149 &    $43.62\pm0.04~$     & $1494\pm 802\phn$   & $1196\pm141$          &  1 \\
\nodata            &   \nodata   & 49752-49875 &    $43.69\pm0.04~$     & $1381\pm 528\phn$   & $1115\pm103$          &  1 \\
\nodata            &   \nodata   & 50011-50262 &    $43.47\pm0.05~$     & $1521\pm59\phn\phn$ & $\phn788\pm29\phn$    &  1  \\
Mrk 142            &   0.04494   & 54506-54618 &    $43.54\pm0.04~$     & $1368\pm 379\phn$   & $\phn859\pm102$       &  5, 6 \\
Mrk 202            &   0.02102   & 54505-54617 &    $42.20\pm0.18~$     & $1354\pm 250\phn$   & $\phn659\pm 65\phn$   &  5, 6\\
Mrk 279            &   0.03045   & 50095-50289 &    $43.64\pm0.08~$     & $3385\pm 349\phn$   & $1420\pm 96\phn$      &  9\\
Mrk 290            &   0.02958   & 54180-54321 &    $43.11\pm0.06~$     & $4270\pm 157\phn$   & $1609\pm 47\phn$      &  10 \\
Mrk 335            &   0.02578   & 48156-49338 &    $43.70\pm0.06~$     & $1629\pm145\phn$    & $\phn917\pm 52\phn$   &  1 \\
\nodata            &   \nodata   & 49889-50118 &    $43.78\pm0.05~$     & $1375\pm357\phn$    & $\phn948\pm113$       &  1 \\
\nodata            &   \nodata   & 55431-55569 &    $43.68\pm0.06~$     & $1025\pm35\phn\phn$ & $1293\pm 64\phn$      &  2 \\ 
Mrk 509            &   0.03440   & 47653-50374 &    $44.13\pm0.05~$     & $2715\pm 101\phn$   & $1276\pm 28\phn$      &  1 \\
Mrk 590            &   0.02638   & 48090-48323 &    $43.53\pm0.07~$     & $1675\pm 587\phn$   & $\phn789\pm 74\phn$   &  1 \\
\nodata            &   \nodata   & 48848-49048 &    $43.07\pm0.11~$     & $2566\pm 106\phn$   & $1935\pm 52\phn$      &  1 \\
\nodata            &   \nodata   & 49183-49338 &    $43.32\pm0.08~$     & $2115\pm 575\phn$   & $1251\pm 72\phn$      &  1 \\
\nodata            &   \nodata   & 49958-50112 &    $43.59\pm0.06~$     & $1979\pm 386\phn$   & $1201\pm130$          &  1 \\
Mrk 817            &   0.03145   & 49000-49212 &    $43.73\pm0.05~$     & $3515\pm 393\phn$   & $1392\pm 78\phn$      &  1 \\
\nodata            &   \nodata   & 49404-49528 &    $43.61\pm0.05~$     & $4952\pm 537\phn$   & $1971\pm 96\phn$      &  1\\
\nodata            &   \nodata   & 49752-49924 &    $43.61\pm0.05~$     & $3752\pm 995\phn$   & $1729\pm158$          &  1\\
\nodata            &   \nodata   & 54200-54331 &    $43.78\pm0.05~$     & $5627\pm  30\phn$   & $2025\pm  5\phn\phn$  &  10\\
Mrk 1310           &   0.01956   & 54516-54618 &    $42.23\pm0.17~$     & $1602\pm 250\phn$   & $\phn755\pm138$       &  5, 6\\
NGC 3227           &   0.00386   & 51480-54273 &    $42.24\pm0.11~$     & $3578\pm83\phn\phn$ & $1376\pm 44\phn$      &  10 \\
NGC 3516           &   0.00884   & 54181-54300 &    $42.73\pm0.21~$     & $5175\pm96\phn\phn$ & $1591\pm 10\phn$      &  10 \\
NGC 3783           &   0.00973   & 48607-48833 &    $42.55\pm0.18~$     & $3093\pm 529\phn$   & $1753\pm141$          &  11\\
NGC 4051           &   0.00234   & 54180-54311 &    $41.96\pm0.20~$     & $1034\pm41\phn\phn$ & $\phn927\pm64\phn$    &  12\\
NGC 4151           &   0.00332   & 53430-53471 &    $42.09\pm0.22~$     & $4711\pm 750\phn$   & $2680\pm 64\phn$      &  13\\
NGC 4253           &   0.01293   & 54509-54618 &    $42.51\pm0.13~$     & $\phn834\pm1260$    & $\phn516\pm 91\phn$   &  5, 6\\
NGC 4593           &   0.00900   & 53391-53579 &    $42.87\pm0.18~$     & $4141\pm 416\phn$   & $1561\pm55\phn$       &  14\\
NGC 4748           &   0.01463   & 54505-54618 &    $42.49\pm0.13~$     & $1212\pm 173\phn$   & $\phn657\pm91\phn$    &  5, 6\\
NGC 5548           &   0.01717   & 47509-47809 &    $43.33\pm0.10~$     & $4044\pm 199\phn$   & $1687\pm 56\phn$      &  15\\
\nodata            &   \nodata   & 47861-48179 &    $43.08\pm0.11~$     & $4664\pm 324\phn$   & $1882\pm 83\phn$      &  15\\
\nodata            &   \nodata   & 48225-48534 &    $43.29\pm0.10~$     & $5776\pm 237\phn$   & $2075\pm 81\phn$      &  15\\
\nodata            &   \nodata   & 48623-48898 &    $43.01\pm0.11~$     & $5691\pm 164\phn$   & $2264\pm 88\phn$      &  15\\
\nodata            &   \nodata   & 48954-49255 &    $43.26\pm0.10~$     & $2543\pm 605\phn$   & $1909\pm129$          &  15\\
\nodata            &   \nodata   & 49309-49636 &    $43.32\pm0.10~$     & $7202\pm 392\phn$   & $2895\pm114$          &  15\\
\nodata            &   \nodata   & 49679-50008 &    $43.46\pm0.09~$     & $6142\pm 289\phn$   & $2247\pm134$          &  15\\
\nodata            &   \nodata   & 50044-50373 &    $43.37\pm0.09~$     & $5706\pm 357\phn$   & $2026\pm 68\phn$      &  15\\
\nodata            &   \nodata   & 50434-50729 &    $43.18\pm0.10~$     & $5541\pm 354\phn$   & $1923\pm 62\phn$      &  15\\
\nodata            &   \nodata   & 50775-51085 &    $43.52\pm0.09~$     & $4596\pm 505\phn$   & $1732\pm 76\phn$      &  15\\
\nodata            &   \nodata   & 51142-51456 &    $43.44\pm0.09~$     & $6377\pm 147\phn$   & $1980\pm 30\phn$      &  15\\
\nodata            &   \nodata   & 51517-51791 &    $43.05\pm0.11~$     & $5957\pm 224\phn$   & $1969\pm 48\phn$      &  15\\
\nodata            &   \nodata   & 51878-52174 &    $43.05\pm0.11~$     & $6247\pm 343\phn$   & $2173\pm 89\phn$      &  15\\
\nodata            &   \nodata   & 53431-53471 &    $42.90\pm0.13~$     & $8047\pm1268$       & $3078\pm197$          &  16\\
\nodata            &   \nodata   & 54508-54618 &    $42.95\pm0.11~$     &$11177\pm2266$       & $4270\pm292$          &  5, 6\\
\nodata            &   \nodata   & 54180-54332 &    $42.93\pm0.12~$     & $4849\pm 112\phn$   & $1822\pm 35\phn$      &  10\\
NGC 6814           &   0.00521   & 54546-54618 &    $42.05\pm0.29~$     & $3277\pm 297\phn$   & $1610\pm108$          &  5, 6 \\
PG 0026+129        &   0.14200   & 48545-51084 &    $44.91\pm0.02~$     & $1719\pm495\phn$    & $1773\pm285$          &  17 \\  
PG 0052+251        &   0.15500   & 48461-51084 &    $44.75\pm0.03~$     & $4615\pm 381\phn$   & $1783\pm 86\phn$      &  17 \\
PG 0804+761        &   0.10000   & 48319-51085 &    $44.85\pm0.02~$     & $2012\pm 845\phn$   & $1971\pm105$          &  17 \\  
PG 0953+414        &   0.23410   & 48319-50997 &    $45.13\pm0.01~$     & $3002\pm 398\phn$   & $1306\pm144$          &  17 \\
PG 1226+023        &   0.15834   & 48361-50997 &    $45.90\pm0.02~$     & $2598\pm 299\phn$   & $1777\pm150$          &  17 \\
PG 1229+204        &   0.06301   & 48319-50997 &    $43.64\pm0.06~$     & $3415\pm 320\phn$   & $1385\pm111$          &  17 \\
PG 1307+085        &   0.15500   & 48319-51042 &    $44.79\pm0.02~$     & $5058\pm 524\phn$   & $1820\pm122$          &  17 \\
PG 1411+442        &   0.08960   & 48319-51038 &    $44.50\pm0.02~$     & $2398\pm 353\phn$   & $1607\pm169\phn$      &  17 \\
PG 1426+015        &   0.08647   & 48334-51042 &    $44.57\pm0.02~$     & $6323\pm1295$       & $3442\pm308$          &  17 \\
PG 1613+658        &   0.12900   & 48397-51073 &    $44.71\pm0.03~$     & $7897\pm1792$       & $2547\pm342$          &  17 \\
PG 1617+175        &   0.11244   & 48362-51085 &    $44.33\pm0.02~$     & $4718\pm 991\phn$   & $2626\pm211$          &  17 \\
PG 1700+518        &   0.29200   & 48378-51084 &    $45.53\pm0.03~$     & $1846\pm 682\phn$   & $1700\pm123$          &  17 \\
PG 2130+099        &   0.06298   & 55430-55557 &    $44.14\pm0.03~$     & $2097\pm 102\phn$   & $1825\pm 65\phn$      &  2\\
SBS 1116+583A      &   0.02787   & 54505-54618 &    $42.07\pm0.28~$     & $3604\pm1123$       & $1528\pm184$          &  5, 6
\enddata
\tablerefs{(1) \cite{Peterson1998};  (2) \cite{Grier2012}; (3) \cite{Dietrich1998}; (4) \cite{Dietrich2012};
(5) \cite{Bentz2009b}; (6) \cite{Walsh2009}; (7) \cite{Rodriguez1997}; (8) \cite{Santos1997}; 
(9) \cite{Santos2001}; (10) \cite{Denney2010}; (11) \cite{Onken2002}; (12) \cite{Denney2009}; 
(13) \cite{Bentz2006}; (14) \cite{Denney2006}; (15) \cite{Peterson2002} and references therein;
(16) \cite{Bentz2007}; (17) \cite{Kaspi2000}.}
\label{tab_obj}
\end{deluxetable}

\newpage
\LongTables
\begin{deluxetable}{lccccccc}
\tablecolumns{8}
\tablewidth{1.0\textwidth}
\tablecaption{Parameter Values for BLR Modeling.}
\tablehead{
  \colhead{Object} 
& \colhead{$R_{\rm BLR}$}    
& \colhead{$\theta_{\rm inc}$}
& \colhead{$\theta_{\rm opn}$}
& \colhead{$\gamma$}
& \colhead{$\log\tau_{\rm d}$}
& \colhead{$f_{\rm BLR}$$^{\rm a}$}
& \colhead{Detrending$^{\rm b}$} \\
  \colhead{}
& \colhead{(light day)}
& \colhead{(degree)}
& \colhead{(degree)}
& \colhead{}
& \colhead{(day)}
& \colhead{}
& \colhead{}
}
\startdata
         3C 120  &   $    51.7\pm    7.5$  &   $      43\pm     22$  &   $      47\pm     24$  &   $   -0.58\pm   0.03$  &   $    2.34\pm   0.31$   &   \nodata  &   Y \\
        \nodata  &   $    25.4\pm    1.2$  &   $      49\pm     21$  &   $      50\pm     24$  &   $   -0.33\pm   0.07$  &   $    1.66\pm   0.44$   &   \nodata  &   \nodata \\
       3C 390.3  &   $    19.8\pm    3.3$  &   $      42\pm     25$  &   $      46\pm     23$  &   $   -0.32\pm   0.04$  &   $    3.08\pm   0.42$   &   \nodata  &   \nodata \\
        \nodata  &   $    41.0\pm    4.1$  &   $       8\pm      6$  &   $      12\pm      9$  &   $   -0.06\pm   0.16$  &   $    2.12\pm   0.51$   &    15.4  &   \nodata \\
        Ark 120  &   $    36.6\pm    6.3$  &   $      44\pm     23$  &   $      47\pm     24$  &   $    0.47\pm   0.22$  &   $    1.83\pm   0.69$   &   \nodata  &   \nodata \\
        \nodata  &   $    30.8\pm   10.2$  &   $      35\pm     25$  &   $      40\pm     23$  &   $   -0.65\pm   0.11$  &   $    2.04\pm   0.51$   &     1.3  &   \nodata \\
        Arp 151  &   $     4.0\pm    0.4$  &   $      29\pm     18$  &   $      36\pm     22$  &   $    0.30\pm   0.04$  &   $    1.88\pm   0.59$   &     1.8  &   \nodata \\
      Fairall 9  &   $    38.6\pm    5.7$  &   $      43\pm     19$  &   $      49\pm     24$  &   $   -0.66\pm   0.02$  &   $    2.30\pm   0.45$   &   \nodata  &   \nodata \\
         Mrk 79  &   $    20.5\pm    7.0$  &   $      42\pm     23$  &   $      45\pm     24$  &   $   -0.22\pm   0.16$  &   $    2.29\pm   0.55$   &   \nodata  &   \nodata \\
        \nodata  &   $    21.2\pm    3.9$  &   $      59\pm     21$  &   $      43\pm     24$  &   $   -0.05\pm   0.12$  &   $    1.88\pm   0.47$   &   \nodata  &   \nodata \\
        \nodata  &   $    20.0\pm    4.2$  &   $      24\pm     21$  &   $      34\pm     21$  &   $   -0.43\pm   0.07$  &   $    2.28\pm   0.49$   &     2.1  &   \nodata \\
        Mrk 110  &   $    22.7\pm    6.4$  &   $      32\pm     21$  &   $      42\pm     22$  &   $   -0.57\pm   0.10$  &   $    0.73\pm   0.68$   &   \nodata  &   \nodata \\
        \nodata  &   $    22.7\pm    4.6$  &   $      36\pm     24$  &   $      47\pm     24$  &   $   -0.19\pm   0.17$  &   $    1.18\pm   0.38$   &   \nodata  &   Y \\
        \nodata  &   $    23.8\pm    2.4$  &   $      14\pm      9$  &   $      22\pm     13$  &   $   -0.41\pm   0.02$  &   $    2.68\pm   0.44$   &     5.1  &   \nodata \\
        Mrk 142  &   $    15.3\pm    2.7$  &   $      68\pm     16$  &   $      45\pm     24$  &   $    2.51\pm   0.27$  &   $    0.82\pm   0.36$   &   \nodata  &   \nodata \\
        Mrk 202  &   $     4.9\pm    1.4$  &   $      53\pm     22$  &   $      51\pm     23$  &   $    3.03\pm   0.31$  &   $    1.04\pm   0.42$   &   \nodata  &   \nodata \\
        Mrk 279  &   $    18.4\pm    4.7$  &   $      15\pm     13$  &   $      20\pm     16$  &   $   -0.38\pm   0.05$  &   $    2.38\pm   0.54$   &     5.3  &   \nodata \\
        Mrk 290  &   $     9.5\pm    4.8$  &   $      24\pm     12$  &   $      33\pm     18$  &   $   -0.61\pm   0.06$  &   $    2.15\pm   0.54$   &     2.2  &   \nodata \\
        Mrk 335  &   $    15.9\pm    2.5$  &   $      51\pm     24$  &   $      47\pm     24$  &   $   -0.25\pm   0.11$  &   $    1.87\pm   0.52$   &   \nodata  &   \nodata \\
        \nodata  &   $    11.1\pm    5.0$  &   $      42\pm     23$  &   $      45\pm     24$  &   $   -0.59\pm   0.10$  &   $    1.45\pm   0.48$   &   \nodata  &   \nodata \\
        \nodata  &   $    15.8\pm    0.8$  &   $      15\pm      8$  &   $      24\pm     11$  &   $   -0.29\pm   0.02$  &   $    2.28\pm   0.45$   &     4.4  &   \nodata \\
        Mrk 509  &   $   108.9\pm    4.0$  &   $      68\pm     11$  &   $      34\pm     21$  &   $   -0.19\pm   0.03$  &   $    2.46\pm   0.27$   &     0.8  &   \nodata \\
        Mrk 590  &   $    23.5\pm    6.2$  &   $      30\pm     19$  &   $      39\pm     26$  &   $    0.21\pm   0.22$  &   $    1.96\pm   0.61$   &     1.5  &   \nodata \\
        \nodata  &   $    16.5\pm    3.6$  &   $      48\pm     24$  &   $      48\pm     24$  &   $    2.25\pm   0.23$  &   $    2.21\pm   0.54$   &   \nodata  &   \nodata \\
        \nodata  &   $    30.3\pm    4.3$  &   $      30\pm     21$  &   $      33\pm     24$  &   $    1.10\pm   0.35$  &   $    1.91\pm   0.55$   &     1.9  &   \nodata \\
        \nodata  &   $    28.8\pm    3.6$  &   $      23\pm     18$  &   $      32\pm     23$  &   $   -0.06\pm   0.14$  &   $    2.24\pm   0.53$   &     2.3  &   \nodata \\
        Mrk 817  &   $    19.4\pm    2.6$  &   $      15\pm     13$  &   $      26\pm     19$  &   $   -0.03\pm   0.20$  &   $    1.27\pm   0.42$   &     4.0  &   Y \\
        \nodata  &   $    19.0\pm    2.7$  &   $      41\pm     22$  &   $      45\pm     23$  &   $    0.14\pm   0.16$  &   $    1.92\pm   0.59$   &   \nodata  &   \nodata \\
        \nodata  &   $    36.1\pm    9.1$  &   $      37\pm     24$  &   $      39\pm     26$  &   $    0.16\pm   0.22$  &   $    1.85\pm   0.48$   &     1.3  &   \nodata \\
        \nodata  &   $    10.7\pm    1.2$  &   $      57\pm     21$  &   $      47\pm     24$  &   $    0.25\pm   0.15$  &   $    1.67\pm   0.29$   &   \nodata  &   Y \\
       Mrk 1310  &   $     4.3\pm    0.5$  &   $      11\pm      9$  &   $      18\pm     14$  &   $    0.69\pm   0.12$  &   $    0.95\pm   0.18$   &     7.6  &   Y \\       
       NGC 3227  &   $     5.1\pm    0.7$  &   $      60\pm     22$  &   $      48\pm     24$  &   $    0.27\pm   0.11$  &   $    0.42\pm   0.13$   &   \nodata  &   Y \\
       NGC 3516  &   $    15.3\pm    1.2$  &   $      61\pm     13$  &   $      36\pm     24$  &   $   -0.25\pm   0.05$  &   $    2.17\pm   0.57$   &     0.9  &   \nodata \\
       NGC 3783  &   $     5.8\pm    3.6$  &   $      43\pm     24$  &   $      47\pm     24$  &   $   -0.07\pm   0.13$  &   $    1.81\pm   0.58$   &   \nodata  &   \nodata \\
       NGC 4051  &   $     2.5\pm    0.7$  &   $      73\pm     14$  &   $      36\pm     24$  &   $   -0.01\pm   0.06$  &   $    0.82\pm   0.20$   &     0.8  &   \nodata \\
       NGC 4151  &   $     7.9\pm    0.9$  &   $      53\pm     18$  &   $      57\pm     21$  &   $    0.22\pm   0.05$  &   $    1.89\pm   0.62$   &   \nodata  &   \nodata \\
       NGC 4253  &   $     7.1\pm    1.6$  &   $      43\pm     23$  &   $      47\pm     23$  &   $    0.41\pm   0.16$  &   $    0.09\pm   0.20$   &   \nodata  &   \nodata \\
       NGC 4593  &   $     4.3\pm    0.7$  &   $      36\pm     24$  &   $      42\pm     23$  &   $   -0.18\pm   0.12$  &   $    1.13\pm   0.39$   &   \nodata  &   \nodata \\
       NGC 4748  &   $     7.3\pm    1.0$  &   $      58\pm     21$  &   $      50\pm     23$  &   $    0.43\pm   0.11$  &   $    0.79\pm   0.27$   &   \nodata  &   \nodata \\
       NGC 5548  &   $    21.9\pm    0.8$  &   $       9\pm      6$  &   $      15\pm      8$  &   $   -0.24\pm   0.03$  &   $    1.92\pm   0.40$   &    11.2  &   \nodata \\
        \nodata  &   $    18.8\pm    1.1$  &   $      19\pm     12$  &   $      29\pm     18$  &   $    0.48\pm   0.05$  &   $    2.57\pm   0.51$   &     3.0  &   \nodata \\
        \nodata  &   $    16.7\pm    1.6$  &   $      29\pm     19$  &   $      38\pm     23$  &   $    0.23\pm   0.12$  &   $    1.52\pm   0.41$   &     1.6  &   \nodata \\
        \nodata  &   $    12.9\pm    1.2$  &   $      62\pm     19$  &   $      44\pm     25$  &   $    1.07\pm   0.07$  &   $    2.46\pm   0.45$   &   \nodata  &   Y \\
        \nodata  &   $    14.3\pm    0.8$  &   $      13\pm      8$  &   $      19\pm     11$  &   $   -0.31\pm   0.05$  &   $    1.66\pm   0.31$   &     6.3  &   Y \\
        \nodata  &   $    15.2\pm    1.4$  &   $      63\pm     16$  &   $      46\pm     24$  &   $    0.21\pm   0.04$  &   $    2.05\pm   0.43$   &   \nodata  &   \nodata \\
        \nodata  &   $    19.9\pm    1.4$  &   $      55\pm     20$  &   $      47\pm     25$  &   $   -0.09\pm   0.05$  &   $    2.17\pm   0.49$   &   \nodata  &   \nodata \\
        \nodata  &   $    16.6\pm    0.8$  &   $      17\pm      9$  &   $      23\pm     12$  &   $   -0.25\pm   0.03$  &   $    2.47\pm   0.43$   &     4.2  &   \nodata \\
        \nodata  &   $    20.4\pm    1.6$  &   $      47\pm     21$  &   $      53\pm     24$  &   $    0.19\pm   0.07$  &   $    1.38\pm   0.38$   &   \nodata  &   \nodata \\
        \nodata  &   $    29.4\pm    1.5$  &   $      42\pm     18$  &   $      59\pm     20$  &   $   -0.20\pm   0.04$  &   $    2.00\pm   0.40$   &   \nodata  &   \nodata \\
        \nodata  &   $    24.4\pm    2.9$  &   $      48\pm     21$  &   $      50\pm     23$  &   $   -0.39\pm   0.05$  &   $    1.54\pm   0.38$   &   \nodata  &   \nodata \\
        \nodata  &   $     8.2\pm    2.1$  &   $      61\pm     21$  &   $      43\pm     24$  &   $    0.15\pm   0.11$  &   $    0.87\pm   0.31$   &   \nodata  &   \nodata \\
        \nodata  &   $    19.5\pm    2.4$  &   $      72\pm     15$  &   $      33\pm     22$  &   $    0.89\pm   0.15$  &   $    1.20\pm   0.30$   &     0.8  &   \nodata \\
        \nodata  &   $     5.9\pm    1.3$  &   $      40\pm     23$  &   $      46\pm     23$  &   $    0.94\pm   0.25$  &   $    1.05\pm   0.74$   &   \nodata  &   \nodata \\
        \nodata  &   $     5.3\pm    0.9$  &   $      65\pm     20$  &   $      43\pm     25$  &   $    0.08\pm   0.07$  &   $    1.56\pm   0.60$   &   \nodata  &   \nodata \\
        \nodata  &   $     3.9\pm    0.9$  &   $      54\pm     23$  &   $      48\pm     24$  &   $    1.46\pm   0.11$  &   $    2.13\pm   0.56$   &   \nodata  &   Y \\
       NGC 6814  &   $     6.9\pm    0.7$  &   $      56\pm     21$  &   $      55\pm     23$  &   $   -0.03\pm   0.08$  &   $    0.95\pm   0.34$   &   \nodata  &   Y \\
    PG 0026+129  &   $   136.4\pm   32.2$  &   $      56\pm     21$  &   $      48\pm     24$  &   $   -0.41\pm   0.05$  &   $    2.37\pm   0.27$   &   \nodata  &   \nodata \\
    PG 0052+251  &   $   104.1\pm    8.0$  &   $      56\pm     20$  &   $      44\pm     26$  &   $   -0.24\pm   0.06$  &   $    2.45\pm   0.30$   &   \nodata  &   \nodata \\
    PG 0804+761  &   $   109.4\pm    7.5$  &   $      27\pm     12$  &   $      32\pm     20$  &   $   -0.53\pm   0.03$  &   $    2.54\pm   0.28$   &     2.1  &   Y \\
    PG 0953+414  &   $   143.1\pm   28.6$  &   $      38\pm     20$  &   $      37\pm     23$  &   $   -0.28\pm   0.09$  &   $    2.24\pm   0.29$   &     1.4  &   Y \\
    PG 1226+023  &   $   392.5\pm   45.7$  &   $      58\pm     20$  &   $      45\pm     25$  &   $    0.10\pm   0.12$  &   $    2.67\pm   0.33$   &   \nodata  &   \nodata \\
    PG 1229+204  &   $    36.0\pm    6.6$  &   $      53\pm     23$  &   $      47\pm     24$  &   $    0.36\pm   0.13$  &   $    2.01\pm   0.23$   &   \nodata  &   \nodata \\
    PG 1307+085  &   $   120.6\pm   30.1$  &   $      54\pm     23$  &   $      46\pm     24$  &   $    0.34\pm   0.18$  &   $    1.94\pm   0.22$   &   \nodata  &   \nodata \\
    PG 1411+442  &   $    45.6\pm   12.5$  &   $      36\pm     25$  &   $      48\pm     24$  &   $   -0.44\pm   0.09$  &   $    1.62\pm   0.22$   &   \nodata  &   \nodata \\
    PG 1426+015  &   $   122.4\pm   25.5$  &   $      46\pm     22$  &   $      48\pm     24$  &   $   -0.30\pm   0.09$  &   $    2.84\pm   0.34$   &   \nodata  &   \nodata \\
    PG 1613+658  &   $   137.1\pm    8.1$  &   $      24\pm     10$  &   $      19\pm     15$  &   $   -0.27\pm   0.08$  &   $    2.83\pm   0.33$   &     3.7  &   \nodata \\
    PG 1617+175  &   $   116.4\pm   23.9$  &   $      47\pm     21$  &   $      51\pm     24$  &   $   -0.30\pm   0.07$  &   $    2.48\pm   0.27$   &   \nodata  &   \nodata \\
    PG 1700+518  &   $   372.4\pm   73.2$  &   $      55\pm     23$  &   $      51\pm     24$  &   $   -0.14\pm   0.17$  &   $    2.16\pm   0.29$   &   \nodata  &   Y \\
    PG 2130+099  &   $    32.7\pm    3.0$  &   $      51\pm      5$  &   $       9\pm      8$  &   $   -0.20\pm   0.06$  &   $    2.43\pm   0.59$   &     1.6  &   \nodata \\
  SBS 1116+583A  &   $     3.3\pm    0.8$  &   $      52\pm     23$  &   $      49\pm     23$  &   $    0.56\pm   0.15$  &   $    0.45\pm   0.22$   &   \nodata  &   \nodata
\enddata
\tablenotetext{a}{Only $f_{\rm BLR}$ for $\theta_{\rm opn}<40^\circ$ is calculated using Equation (\ref{eqn_factor}).}
\tablenotetext{b}{``Y'' means detrending is implemented.}
\label{tab_par}
\end{deluxetable}

\end{document}